\documentclass[12pt,preprint]{aastex}

\begin{document}

\title{The anomalous accretion disk of the Cataclysmic Variable RW Sextantis \footnotemark[1]}
\footnotetext[1]
{Based on observations made with the NASA/ESA Hubble Space Telescope, obtained at the
Space Telescope Science Institute, which is operated by the Association of Universities 
for Research in Astronomy, Inc. under NASA contract NAS5-26555, and the NASA-CNES-CSA
{\it Far Ultraviolet Explorer}, which is operated for NASA by the Johns Hopkins University
under NASA contract NAS5-32985} 


\author{Albert P. Linnell$^2$, Patrick Godon$^3$, Ivan Hubeny$^4$, Edward M. Sion$^5$,
and Paula Szkody$^6$}

\affil{$^2$Department of Astronomy, University of Washington, Box 351580, Seattle,
WA 98195-1580\\
$^3$Department of Astronomy and Astrophysics, Villanova University,
Villanova, PA 19085\\
$^4$Steward Observatory and Department of Astronomy,
University of Arizona, Tucson, AZ 85721\\
$^5$Department of Astronomy and Astrophysics, Villanova University,
Villanova, PA 19085\\
$^6$Department of Astronomy, University of Washington, Box 351580, Seattle,
WA 98195-1580\\
}

\email{$^2$linnell@astro.washington.edu\\
$^3$patrick.godon@villanova.edu\\
$^4$hubeny@as.arizona.edu\\
$^5$edward.sion@villanova.edu\\
$^6$szkody@astro.washington.edu\\
}

\begin{abstract}

 Synthetic spectra covering the wavelength range 900\AA~to 3000\AA~provide an accurate fit, 
 established by a 
 ${\chi}_{\nu}^2$ 
 analysis, to a combined 
 observed spectrum of RW Sextantis. 
 Two separately calibrated distances to the system establish the
 synthetic spectrum comparison on an absolute flux basis but with two alternative scaling factors, requiring 
 alternative values of $\dot{M}$ for final models.
 Based on comparisons for a range of $\dot{M}$ values, 
 the observed spectrum does not follow the standard model. Rather than the exponent 0.25 in the expression
 for the radial temperature profile, a value close to 0.125 produces a synthetic spectrum with an 
 accurate fit to 
 the combined spectrum.	A study of time-series $FUSE$ spectra shows that a proposed warped or tilted disk is not
 supported by the data; an alternative proposal is that an observed non-axisymmetric wind results from an
 interaction with the mass transfer stream debris.

\end{abstract}


\keywords{Stars:Novae,Cataclysmic Variables,Stars:White Dwarfs,Stars:
Individual:Constellation Name: RW Sextantis}


\section{Introduction}

Cataclysmic variables (CVs) are semi-detached binary stars in which a late main sequence star 
loses mass onto a white dwarf (WD) by Roche lobe overflow \citep{w1995}. 
In non-magnetic systems the mass
transfer stream produces an accretion disk with mass transport inward and angular momentum 
transport outward, driven by viscous processes. The accretion disk may extend inward to the WD; the
outer boundary extends to a tidal cutoff limit imposed by the secondary star in the steady state case.
If the mass transfer rate is below a certain limit, the accretion disk is unstable
and undergoes brightness cycles (outbursts), and if above the limit, the 
accretion disk is stable against outbursts \citep{osaki1996}.
The latter objects (which have no recorded outburst of any type) are called nova-like (NL) systems.
As shown by \citet{Cowley1977}, 
\citet{beuermann} (hereafter BEU) and \citet{prinja2003}, RW Sex is 
a NL system (see \citet{w1995} and \citet{lasota2001} for a detailed discussion).
RW Sex has a $Hipparcos$~\citep{perry1997} parallax and a separately calibrated distance (BEU),
of importance in constraining the mass transfer rate $\dot{M}$. BEU provide estimates of 
the component stellar
masses and the orbital inclination; \citet{gstein} provide an estimate of 
$\dot{M}=1.0{\times}10^{-8}M_{\odot}~{\rm yr}^{-1}$ while \citet{wade1988} quotes a value of
$\dot{M}=3.0{\times}10^{-9}M_{\odot}~{\rm yr}^{-1}$ from \citet{patterson}.
A $FUSE$ spectrum of RW Sex is available as well as HST and $IUE$ spectra. $FUSE$ spectra are
important in constraining the WD $T_{\rm eff}$ as well as $\dot{M}$ since the peaks of the radiation
curves fall in the $FUSE$ spectral range. 

NL systems are of special interest because they are expected to have an accretion disk radial temperature
profile	given by an analytic expression \citep[eq.5.41]{fr92} (hereafter FKR, cf eq.~(2) below) which defines
the so-called standard model; the expression includes the mass transfer rate $\dot{M}$ as an
explicit variable. 
In NL systems which are above the period gap \citep{howell2001} 
the accretion disk
dominates the system spectrum (with the exception of BK Lyn
\citep{Zellem2009}). 
Fitting a synthetic spectrum based on the analytic model (a proxy of
the accretion disk temperature profile) to an
observed spectrum potentially determines $\dot{M}$. This physical parameter is of basic importance
since it controls the evolution of CV systems \citep{howell2001}. But the  analytic expression
also is an explicit function of the WD mass, $M_{\rm wd}$, and the mass must be determined independently.
\citet{wade1988}, using $IUE$ spectra, showed that NL systems systematically disagree with 
the standard model when either black body spectra or
Kurucz stellar model spectra are used to represent the accretion disk. The Kurucz spectra were too "blue",
i.e., had too large a spectral gradient as compared with the $IUE$ spectra and this was specifically
true of RW Sex. Wade concluded that
the problem was with use of Kurucz spectra; they are not suited to representation of accretion disks.
\citet{hubeny1990} has developed a procedure for modeling annuli of accretion disks that explicitly
includes calculation of synthetic spectra using the standard model. In this paper we apply the Hubeny
model to determine system parameters for RW Sex and to test whether the system conforms to the standard model.

\section{The $FUSE$ and HST/GHRS Spectra}

Table~1 presents the observations log. This is the same observation set studied by \citet{prinja2003}
and discussed by those authors.

\subsection{The $FUSE$ Spectrum}

The $FUSE$ spectrum of RW Sex consists of 25 exposures (spacecraft 
orbits) totaling more than 25ks of calibrated (good) exposure time.
The	$FUSE$ spectrum (flux and errors) were extracted using the CalFUSE software, initially binned at
0.016\AA. The data were rebinned
at an interval of 0.5\AA~using the IRAF command trebin. 
Typical noise levels are of order 3\% of the signal level.
Each spectrum consists of a list of flux values (${\rm erg}~{\rm cm}^{-2}~{\rm s}^{-1}~{\rm \AA}^{-1}$)
and associated noise values in the same units. 
The $FUSE$ spectrum covers the interval 903\AA~to 1188\AA.

In Figure~1 we present the combined spectrum
(25ks) annotated with absorption lines.  
It is impractical to place individual error bars on the plotted points.
Besides the obvious 
Ly$\beta$ and Ly$\gamma$ broad features (due to the disk and
accreting WD), the system exhibits broad absorption
lines (which appear blue shifted in this composite by as much 
as 3\AA) and a multitude of very sharp absorption lines
of interstellar origin. We describe these absorption lines
below and list some of them in Table~2 and Table~3.    
The observed wavelengths of the sharp absorption lines listed in
Table~2 were measured in the 1SiC channel for 
$\lambda < 1100$\AA\ , and 2LiF channel for 
$\lambda > 1100$\AA\ ; as a consequence the observed 
ISM lines are blue shifted by $\sim 0.05$\AA\  for   
$\lambda < 1100$\AA\ and red shifted by about the same
amount for $\lambda > 1100$\AA\ as the channels
are not precisely aligned. When we co-added the channels
and different orbits we aligned all the spectral segments.

NL system spectra typically exhibit strong absorption lines from high excitation ions, a signature
of a wind believed to originate from the accretion disk \citep{pereyra2006,proga2003a,proga2003b}.
See the review by \citet{drew2000}. The presence of these lines is the largest current obstacle to
spectrum synthesis modeling of the accretion disk in NL systems.
In the present instance it is necessary to mask appreciable sections of the $FUSE$ spectrum
because we have no current ability to model the wind.
This requirement reduces the number of lines that we fit in the $FUSE$ spectrum.

\subsection{The ISM Hydrogen lines} 

The FUSE spectrum of RW Sex 
exhibits a forest of molecular hydrogen absorption lines 
identified by the Werner (W) and Lyman (L) bands, upper vibrational level
(ranging from 16 to 1) and rotational transition R, P and Q with
lower rotational state J=1,2 and 3. For example, starting at 
933.2\AA\    there is the L16R2 line, and next to it at 933.6\AA\ 
the W4Q3 line. There are more than 100 molecular hydrogen lines
extending to $\sim$1100\AA\ (L0R2 1110.10\AA ).
We have marked these lines (vertical tick mark) in the upper part of each panel 
in Figure~1 (annoted MH). 
See \citet{sembach} for the entire list of molecular hydrogen lines
and their wavelengths. 

The atomic hydrogen lines (Lyman series) are marked below 
each panel.  

The feature at 1152\AA\ is a fixed pattern noise (FPN) due to the
FUSE detectors. 
 
\subsection{The ISM metal lines}
  
In addition to the hydrogen atomic and molecular lines,
we identify many orders of neutral oxygen (O\,{\sc i}) starting 
at 916.9\AA\ (the 26th order) and extending to 
about $\sim 1000$\AA\ with 
O\,{\sc i} 4th and 3rd orders at $\lambda \lambda$1025.9 and $\lambda \lambda$1039.3.  
Not all the lines are listed in Table~2, but lines from all the
orders (between 24u and 3u) are identified. 
In the comments column, letter u stands for 
an ultraviolet multiplet and the preceeding number is the
multiplet number (see \citet{morton2000,morton2003}).   
The oxygen lines
are marked with vertical ticks (at mid-height; annoted OI).
 
The FUSE spectrum of RW Sex is rich in oxygen lines in addition to
metal lines which are frequently observed 
in FUSE spectra of CVs. These additional metal lines are 
Ar\,{\sc i}~($\lambda \lambda$ 1048.2, 1066.7), 
C\,{\sc ii}~($\lambda \lambda$ 1036.34, 1037.1), 
Fe\,{\sc ii} (bottom panel of Figure~1),
N\,{\sc i} ~($\lambda \sim$1034), 
N\,{\sc ii}~ ($\lambda \sim$1084). 
We also 
identify many phosphorus (P\,{\sc ii}) lines between 
$\sim 960$\AA\  and $\sim 965$\AA\  as well as a single
line at 1152.8\AA\ (distinct from the well-known 
fixed pattern noise - FPN - introduced by the FUSE detectors
at 1152\AA\ ; both this P\,{\sc ii} line and the FPN feature
are well documented \citep{sembach}). 
It is not unusual to find that a FUSE spectrum rich in 
ISM O\,{\sc i} lines also exhibits many P\,{\sc ii} lines;
P\,{\sc ii} lines are an excellent neutral oxygen  
tracer in various physical environments in the ISM and are
particularly useful as a proxy for O\,{\sc i} lines when these
are saturated or blended \citep{lebouteiller}.  

\subsection{The HST spectra}

The composite HST spectrum of RW Sex shown in Figure~2 
is a combination of two GHRS spectra, z37v0107t and z37v0108t.  
 
The HST spectrum of RW Sex is characterized by deep and broad
absorption lines from the CV source 
together with much shallower and sharper 
lines from the ISM. The lines are identified in Table~4.  

The broad absorption lines can be divided  into two distinct 
groups. The first group of broad lines includes 
Si\,{\sc iii} ($\sim1140$, $\sim 1205$, $\sim 1300$, \& 1327 \AA ),   
Si\,{\sc iv} ($\sim 1392$ \& 1401 \AA ), 
C\,{\sc iii} ($\sim 1173$ \AA ), and He\,{\sc ii} ($\sim 1638$ \AA ) lines,
all blue-shifted by $\sim$1.8-2.5\AA~with a width of
about 5-8\AA. The second group of broad lines includes 
N\,{\sc v} ($\sim 1234$ \& 1237 \AA ) and C\,{\sc iv} ($\sim 1543$ \AA )  
and is distinct from the first group in that the N and C lines are blue shifted
by as much as 5\AA~and 6\AA~respectively. The lines also appear to be
broader with a width of more than 10\AA. It is likely that the
N\,{\sc v} and C\,{\sc iv} lines form in a hotter region expanding
faster than the lines from the first group.   
The Ly$\alpha$ line stands apart in that it is blue-shifted
by less then 1\AA.  

The sharp absorption lines all have a width of about
$\sim 0.5$\AA~or less and are blue-shifted by only 0.2-0.5\AA. 
The lines are listed in Table~4 and we tentatively identify,
as in the FUSE spectrum, some P\,{\sc ii} phosporus lines 
at 1142.5, 1152.6 (also detected in the FUSE spectrum), 
and 1301.7 \AA~(which could be affected
by the O\,{\sc i} 1302 line).

\section{The $IUE$ spectra}

Table~5 lists the low dispersion $IUE$ spectra. There are three pairs of spectra covering the
range 1150\AA~to 3350\AA. We designate them as case1 (SWP01671+LWR01583), case2 (SWP02494+LWR03071), 
and case3 (SWP07500+LWR06494) in temporal sequence.
Separately, the SWP spectra cover the range 1150\AA~to 1978\AA~and the LWR spectra cover the range
1851\AA~to 3350\AA, tabulated in
${\rm erg}~{\rm cm}^{-2}~{\rm s}^{-1}~{\rm \AA}^{-1}$. 
The tabular interval of both spectra is about 2\AA~and each has an associated tabulated noise
level in the same units as the flux. As with the $FUSE$ spectrum, typical noise levels are 3\% of the signal.
We used the IUEDAC software package to extract the spectra, with zero correction for reddening,
and including the $\sigma$ value at each tabular wavelength,
and processed the spectra with the standard recalibration procedure \citep{massa}.
We examined the overlap region of each SWP and LWR pair and found that the case1 pair fitted very
well without scaling either spectrum. 

\section{Merging $FUSE$, HST and $IUE$ spectra}

Our model simulations concern fitting the observed continuum. The presence of broad absorption lines
of high excitation species in the $FUSE$ spectrum, which our model does not simulate, requires masking
of parts of the $FUSE$ spectrum and so reduces the utility of the $FUSE$ spectrum in choosing among
possible models. In addition, the short spectral range of the $FUSE$ spectrum makes a stringent test
of the spectral gradient fit more difficult. However, the peaks of the emission profiles of both the WD
and the accretion disk fall in the $FUSE$ range and the sensitivity to different WD $T_{\rm eff}$ values
and accretion disk $\dot{M}$ values is much greater than in HST or $IUE$ spectra. In view of those
features, we combine the $FUSE$ and $IUE$ spectra to obtain the best constraint on model parameters. The issue
now is to choose the optimum $IUE$ spectrum. We do not include the HST spectra in our simulations because of
the restricted wavelength range of those spectra. 

Figure~3 shows the fit of the various spectra. Identifications of the spectra are in the figure legend.
The mean of HST z37v0108t and z37v0109t (Table~1) is not plotted; it is similar to the mean of z37v0104t plus z37v0105t.
In this overlap region the $IUE$ case1 (orange line) agrees well with the HST (blue) spectra. The other two
$IUE$ spectra are roughly accordant, with the exception of the case3 L$\alpha$ geocoronal $\lambda1220$ line.
The case-to-case variation of the $IUE$ spectra indicates some temporal variation that is most marked in
the absorption lines.

The black line is the $FUSE$ spectrum as observed; it is discrepant from both the HST and $IUE$ spectra.
Two possible explanations of the discrepancy are: (1) there was an error in the reduction of the spectrum,
and, (2) RW Sex is slightly variable and was at a lower luminosity level at the time of the $FUSE$ observation.
We have checked the reduction procedure and find no error. The MAST preview and the \citet{prinja2003} paper agree with
our data tabulation.
The mutual agreement of the HST and $IUE$ spectra support the postulate that the $FUSE$ spectrum was
anomalously weak at the time of observation. 
The red line is the $FUSE$ spectrum divided by 0.82; the fit to the HST spectra now is excellent and,
within the noise level of the $IUE$ spectra, the fit to the $IUE$ spectra is good.

We deleted the region of $IUE$ case1 that overlaps the $FUSE$ spectrum and added 
the remaining part of case1 to the $FUSE$ spectrum divided by 0.82. This combined spectrum shows strong absorption lines
associated with a wind/chromosphere (Figure~1) and other features like the P Cygni CIV line at 1540\AA~(Figure~2).
The synthetic spectrum does not model those features and it is necessary to mask them.
At the same time, the widths of strong absorption lines are discriminants among different models. We were careful
to mask features that clearly arose from a wind/chromosphere while minimizing the total amount of masking.

The $FUSE$ data we simulate consists of flux
values, $F_i$ (${\rm erg}~{\rm cm}^{-2}~{\rm s}^{-1}~{\rm \AA}^{-1})$, and associated noise values, $\sigma_{\rm i}$. 
Calculation of the reduced 
${\chi}_{\nu}^2$
follows from the equation
\begin{equation}
{\chi}_{\nu}^2 = 1/(N-M)\sum_{i=1}^N\Big(\frac{F_i-f({\lambda}_i)}{{\sigma}_i}\Big)^2
\end{equation}
where N is the number of observed wavelengths, M is the number of model parameters (identified and discussed
below),
$F_i$ is the observed flux at a particular wavelength, $\sigma_i$
is the associated uncertainty of $F_i$, and $f({\lambda}_i)$ is the flux calculated by the model.
Our model consists of calculated flux values at a tabular interval of 0.1\AA; the fitting process interpolates
among the calculated flux values to the exact tabular wavelength of each observed data point.

A rough test of a good model fit to an observed data set is a 
${\chi}_{\nu}^2 {\sim} 1$ \citep[ch.12.3]{taylor}.  
In our case,
a final ${\chi}_{\nu}^2 {\sim} 1$ would indicate essentially a perfect fit at the noise level--an accuracy
unachievable in the presence of unmodeled spectral features. 
However, if we do use the available noise level
tabulation, the calculated ${\chi}_{\nu}^2$ variation from model to model still provides a measure of 
relative model-to-model quality of fit.	The subsequent discussion, \S8., includes a more detailed test.
 
\section{ Initially adopted system parameters}

By the basic paradigm of CV evolution \citep{howell2001} there is a close relation between
orbital period and donor mass. In a recent paper, \citet{knigge2007} has calibrated the
relations among $P$, $M_2$, $R_2$, $T_{\rm eff,2}$, and donor spectral type. Thus, given an
orbital period below the tabular upper limit of $6^{\rm h}$, $M_2$ can be determined independently 
of other parameters. If there is an observationally-determined mass ratio,
the WD mass follows directly.

Table~6 (and tablenotes) lists initially adopted parameters and their sources. Several of the system parameters are 
poorly known.
\citet{knigge2006,knigge2007} determines $M_2=0.67M_{\odot}$ for $P=0.24507^{\rm d}$.
BEU determine a mass ratio $1/q=1.35{\pm}0.1$ which, with $M_2$, produces $M_{\rm wd}=0.9$. 
\citet{panei2000} lists a WD radius of $8.82{\times}10^{-3}R_{\odot}$
for a $0.90M_{\odot}$ homogeneous zero temperature Hamada-Salpeter carbon model WD;
in our subsequent study of the observed spectra we correct the radius for the adopted WD $T_{\rm eff}$.

\citet{ver1987} 
determined a preferred value of $E(B-V)=0.0$ with a
rough upper limit of $E(B-V)$ of 0.03, based on a study 
of the 2175\AA~``bump" in $IUE$ spectra. 
\citet{bruch1994} list $E(B-V)=0.02$. \citet{mauche} determine a value of $E(B-V)=0.014$ with appreciable
uncertainty. We adopt $E(B-V)=0.0$.

A number of studies have considered the tidal cutoff boundary, $r_d$, of accretion disks
(\citet{pac1977}; \citet{pp1977}; \citet{wh1988}; \citet{scr1988}; \citet{wk1991}; \citet{g1993}).
These authors agree on $r_d{\sim}0.33D$, where $D$ is the separation of the stellar components.
We adopt this expression for the tidal cutoff radius of the accretion disk.

The $Hipparcos$~\citep{perry1997} parallax, $3.46{\pm}2.44$mas, corresponds to a distance of 289pc. BEU
determine a distance of 150pc, based on an application of Bailey's method \citep{bailey1981} and
different from the~$Hipparcos$~value by slightly more than $1\sigma$. We will
study our results in the context of both distance determinations and we initially
adopt the~$Hipparcos$~value.

With respect to the Table~6 parameters,
the orbital period is known with essentially perfect accuracy as compared to the other parameters.
Of the remaining parameters $M_2$ has no effect on the model spectrum; however $T_{\rm eff}$ of the WD
does affect the model and we include it as an adjustable parameter.

\section{The analysis program: BINSYN}

Our analysis uses the program suite BINSYN \citep{linnell1996}; recent papers \citep{linnell2007,
linnell2008a,linnell2009}
describe its
application to CV systems in detail. 
Briefly, an initial calculation produces a set of annulus
models for a given WD mass, radius, and mass transfer rate. This calculation uses the program
TLUSTY~\footnote{http://nova.astro.umd.edu} \citep{hubeny1988,hl1995}. 
TLUSTY considers the radial and vertical structure of the disk independently;
the radial structure is based on the standard model (FKR) and so follows the prescribed relation
between local $T_{\rm eff}$ and the annulus radius. The vertical structure is solved,
self-consistently, as described by \citet{hubeny1990} and \citet{hh1998}.
The set of annulus models covers the accretion disk from its innermost (WD) radius to 
$r/r_{\rm wd,0}=50.0$, where $r_{\rm wd,0}$ is the radius of the zero-temperature WD. 

The primary source of viscosity in CV accretion disks
is believed to be MRI (magnetorotational instability) \citep{bh1991,b2002}. \citet{hirose2006} 
calculate MHD models with local 
dissipation of turbulence and
show that the vertical extent of an annulus is greater than in previous models.
\citet{blaes2006} show that magnetic support has a significant effect on synthetic spectra of
black hole (BH)
accretion disk annuli in the X-ray region and illustrate the effect in the case of a BH system with 
$M_{\rm BH}=6.62M_{\odot}$ and with an adopted $\alpha=0.02$. There is an insigificant effect in the
visible and UV.	
They are able to simulate the effect of magnetic support within the TLUSTY framework
by adjusting the TLUSTY parameters $\zeta_0$ and $\zeta_1$ and use this simulation to calculate
the effect on synthetic spectra. 
TLUSTY allows for a vertical viscosity profile
within a given annulus; it introduces an assignable division point within an annulus, dividing
the annulus into deep layers and outer layers. 
Within the deep layers, viscosity follows a power law
variation specified by $\zeta_0$, while the outer layers follow a power law variation specified 
by $\zeta_1$. See \citet{hh1998} for details.
The primary effect \citep{blaes2006} of magnetic support is a hardening of the spectrum in the 0.45keV region.
\citet{king2007} point out that MHD models that produce viscosity via MRI require $\alpha$ values
that are a factor of 10 smaller than the $\alpha$=0.1-0.4 required by observational evidence and
suggest that some caution still is needed in accepting the MHD results.	Our models have used the
default values $\zeta_0=\zeta_1=0.0$; this choice produces a constant viscosity vertically within
the annulus.
We have calculated a synthetic spectrum of one annulus that contributes significantly to the system 
synthetic spectrum using the same viscosity parameters used by \citet{blaes2006} and have compared
that synthetic spectrum with our model synthetic spectrum for the same annulus for the spectral
range 900\AA~to 3000\AA. We find no detectable difference, in agreement with \citet{blaes2006}. 
Magnetic support is believed to extend an accretion disk vertically as compared with the standard
model \citep{hirose2006,blaes2006}, an effect confirmed in the case of IX Vel \citep{linnell2007}.
The absence of an eclipse in RW Sex prevents a test in this system.
We assume a fixed $\alpha$ for
the entire accretion disk.

Table~7 lists properties of annuli calculated with TLUSTY(v.203n)
for a mass transfer rate of $5.0{\times}10^{-9}M_{\odot}/{\rm yr}^{-1}$.
This illustrative case is from among the cases we tested in our simulations.
The TLUSTY control file to calculate a given annulus requires a radius of the WD in units of $R_{\odot}$.
All of the annuli used the radius of a zero temperature WD for a homogeneous 
carbon Hamada-Salpeter
$0.90M_{\odot}$ model from \citet{panei2000}.
All of the annuli are solar composition
models, and the models through $r/r_{\rm wd,0}=26.00$ are converged non-LTE models. The remaining models
are so-called grey models (see the TLUSTY Users Guide for an explanation). 
The annulus spectral flux levels vary by a factor of order 4 dex from the innermost annulus to the rim annulus.
The grey models contribute very little to the accretion disk synthetic spectrum and the difference between
the contribution of a grey model synthetic spectrum and a corresponding converged non-LTE model spectrum
is negligible. 
Our adoption of solar composition 
models implicitly assumes that the secondary star,
which supplies the material for the accretion disk, has an atmosphere with a solar composition.

The annulus calculations adopted C,Mg,Al,Si, and Fe as explicit ions in addition to H and He.
Tests show that there are detectable differences between synthetic spectra for only H,He as explicit ions
and models including the metals listed. 
The remaining grey model annuli
use the same set of explicit ions (see the TLUSTY Users Guide for an explanation).
We believe this is the first instance in which non-LTE annulus models using metals as explicit ions
have been used in a CV simulation. As we discuss subsequently, this study performed a ${\chi}_{\nu}^2$ analysis
with $\dot{M}$ values of 1.0, 2.0, 4.0, 5.0, 5.25, 5.5, 5.75, 6.0, 6.25, 6.5 and $7.0{\times}10^{-9}~M_{\odot}~{\rm yr}^{-1}$.
Each value of $\dot{M}$ has an associated table similar to Table~7, for the same range of annulus radii,
with non-LTE models through $r/r_{\rm wd,0}=26.0$ (the limit of the range of convergence) and for the explicit 
ions listed above.

A \citet{ss1973} viscosity parameter $\alpha=0.1$ was used in calculating all annuli.
Each line in Table~7 
represents a separate annulus. Temperatures are in Kelvins.
The column headed by $m_0$ is the column mass, in ${\rm g}~{\rm cm}^{-2}$, 
above the central plane.
The columns headed by $z_0$ and $N_e$ are, respectively, the
height (cm) above the central plane for a
Rosseland optical depth of ${\sim}0.7$ and the electron density (${\rm cm}^{-3}$) at the same level. 
The column headed
by ${\rm log}~g$ is the log gravity (cgs units) in the $z$ direction at a Rosseland optical depth of ${\sim}0.7$. 
The column headed by ${\tau}_{\rm Ross}$
is the Rosseland optical depth at the central plane. We call attention to the fact that the annuli are optically
thick to the outer radius of the accretion disk (and true for the full range of $\dot{M}$ values).

Following the calculation of annulus models for assigned $M_{\rm wd}$ and $\dot{M}$,
program SYNSPEC(v.48)~\footnote{http://nova.astro.umd.edu} \citep{hlj1994} was used to produce a 
synthetic spectrum for each annulus, at a spectral resolution of 0.1\AA.
We adopted solar composition for all synthetic spectra. The synthetic spectra include contributions
from the explicit ions listed above and the remaining first 30 atomic species. 

The ${\chi}_{\nu}^2$ analysis mentioned above needs WD synthetic spectra to pair with each $\dot{M}$
accretion disk model. We calculated non-LTE WD synthetic spectra for $T_{\rm eff}$ values of
35,000K, 40,000K, 45,000K, 50,000K, 55,000K, 60,000K, 65,000K, 70,000K,
and 75,000K with the same list of explicit ions listed above
and at a spectral resolution of 0.1\AA. This WD representation assumes that 
the solar composition
mass transfer stream, after passing through the accretion disk to deposition on the WD, 
so adulterates the WD photosphere that it preserves a solar composition in spite of the
inward diffusion of heavy ions through the WD atmosphere.

BINSYN models the complete CV system, including the WD, secondary star, accretion disk face,
and accretion disk rim as separate entities. The model represents phase-dependent and 
inclination-dependent effects, including eclipse effects and irradiation effects, on all of the system objects.
Calculation of synthetic spectra for the stars requires polar $T_{\rm eff}$ values and gravity-darkening exponents.
We adopted a standard gravity-darkening exponent of 0.25 for the WD; the secondary star makes a negligible
contribution to the system synthetic spectrum, as shown by a comparison of the secondary star synthetic spectrum
and the accretion disk synthetic spectrum, and we do not list the secondary star simulation parameters.
BINSYN represents the accretion disk by a specified
number of annuli (45 in the present case), where that number typically is larger than the number of TLUSTY annulus
models (22 in the present case; see Table~7). 
BINSYN calculates a synthetic spectrum for each of the annuli specified in the BINSYN system model by
interpolation among the array of TLUSTY annulus models, with proper allowance for orbital inclination.
The accretion disk $T_{\rm eff}$ profile may follow the standard model, but, alternatively, the profile
may be specified by a separate input file or by a non-standard radial temperature gradient as discussed below. 

BINSYN has provision to represent a bright spot on the rim face, 
but our present data do not require this facility (\S9).

\section{Standard model simulations of RW Sex}

We adopted system parameters listed in Table~6 for an initial test; the initial value of $\dot{M}$ was
arbitrary and was meant to test whether its value produced a synthetic system spectrum with even an 
approximate fit to the observed spectrum.
On the assumption that the WD contribution would be small, we adopted a WD $T_{\rm eff}=50,000$K based
on our previous studies of CV systems. If necessary, new models could be calculated with a revised
WD $T_{\rm eff}$. The initial choice of $\dot{M}$ ($\dot{M}=1.0{\times}10^{-9}~M_{\odot}~{\rm yr}^{-1}$)
was obviously too small; we successively tried
$\dot{M}=2.0,~4.0,~{\rm and}~5.0{\times}10^{-9}~M_{\odot}~{\rm yr}^{-1}$ and the last choice indicated a value
that might be appropriate but with a problem described below.

In constructing a model accretion disk
it is necessary to allow for the change in WD radius from the
zero temperature model. We used Table~4a of \citet{panei2000} to estimate the 50,000K WD radius at 
$0.0099R_{\odot}$. Our BINSYN model specified 45 annuli for the accretion disk, with assigned
radii and corresponding standard model $T_{\rm eff}$ values listed in Table~8.
The first column in Table~7 and Table~8 specifies inner radii of annuli measured in units of the WD
radius; in Table~7 the unit is a zero temperature WD and in Table~8 the unit is the WD radius for
its assigned $T_{\rm eff}$ of 50,000K, a larger quantity. Thus in Table~8 a given number in column 1 
corresponds to
a larger physical distance, and a lower annulus $T_{\rm eff}$, than for Table~7. 
The synthetic spectrum
for a given BINSYN annulus is calculated by interpolation (temperature-wise) among the Table~7 entries. 
Ideally, we could recalculate the whole series of annulus models (via TLUSTY) based on the new value of 
$R_{\rm wd}$,
but the change in the individual annulus models would be very small and would not be warranted in view
of other uncertainties (e.g., the system parallax).

Table~7 ends
at a $T_{\rm eff}$ of 6970K. The net
effect is that the calculated BINSYN outer annuli have synthetic spectra corresponding to 6970K even
if the specified standard model calls for a lower temperature. The effect on the synthetic spectrum 
of failing to follow the standard model is negligible. Adding
a lower temperature (3500K) stellar synthetic spectrum to the array of Table~7, thereby permitting
interpolation to a lower temperature in the outer BINSYN annuli, made no detectable change
in the system synthetic spectrum. 
As separate issues, the outer annuli temperatures should not fall below 6000K
to avoid outbursts \citep{osaki1996,lasota2001} and tidal dissipation is expected to raise the outer
accretion disk temperature above the standard model value \citep{smak2002}.

The adopted~$Hipparcos$~distance
provided a fixed scaling factor by which to divide the synthetic specrum to superpose it on the
observed ("corrected" $FUSE$ plus $IUE$ case1) spectrum. The spectral gradient of the superposed synthetic 
spectrum was much too
large but the total integrated flux, judged by an eye estimate of the area below the synthetic spectrum, 
appeared about right. We repeated the entire process with $\dot{M}$
values, including the initial test, of 5.0, 5.25. 5.5, 5.75, 6.0, 6.25, 6.5, and 
$7.0{\times}10^{-9}~M_{\odot}~{\rm yr}^{-1}$. In all cases we used a WD $T_{\rm{eff}}$ of 50,000K;
the spectral gradients of all models were too large. Figure~4 shows the results for five models,
identified in the figure legend, and Figure~5 shows the results in the $FUSE$ spectral region.
The usable range of the $IUE$ spectra extend to 3000\AA, matching the spectral interval covered by
our synthetic spectra. 

Smaller mass transfer rates would produce an accretion disk with too small luminosity to fit the observations
(most of the radiation curve would lie below the observed spectrum) and larger mass transfer rates would produce 
an accretion disk with too large luminosity, based on the adopted system distance. 
The contribution of the WD cannot be the source of the too large calculated spectral gradient because its
contribution is much too small; a plot of the accretion disk contribution alone shows nearly identical results.
On a purely empirical basis
we inquire whether an analytic model with a smaller than standard radial temperature gradient but the same total 
luminosity could
fit the observations.

The following equation defines the standard model \citep{fr92}.
\begin{equation}
T_{\rm eff}(R)=\Bigl\{{\frac{3GM\dot{M}}{8{\pi}{R}^3{\sigma}}}\Bigl[1-{\Bigl(\frac{R_{*}}{R}}\Bigr)^{1/2}\Bigr]\Bigr\}^{1/4},
\end{equation}
where $M$ is the mass of the WD, $\dot{M}$ is the mass transfer rate,
and $R_*$ is the radius of the WD. $G$ is the gravitation constant and $\sigma$
is the Stefan-Boltzmann constant.
A smaller radial temperature gradient but the same flux follows from
\begin{equation}
T_{\rm eff}(R)=SCL\Bigl\{{\frac{3GM\dot{M}}{8{\pi}{R}^3{\sigma}}}\Bigl[1-{\Bigl(\frac{R_{*}}{R}}\Bigr)^{1/2}\Bigr]\Bigr\}^{EXP},
\end{equation}
where $EXP$ is a number	smaller than 0.25 and $SCL$ is an empirically-determined number to preserve the standard model
$T_{\rm eff}$ on the left side of the equation (and thereby preserve the same total flux since flux = ${\sigma}T_{\rm eff}^4$).
The significance of this generalization is that $EXP$ becomes an additional model parameter (M in equation~1 now becomes 6).

Equation~3 was included in BINSYN at the time of its development but this is the first instance of its use
in a simulation of a CV system. See \S10 for further discussion.

The calculation proceeds as follows; an initial run at $EXP=0.25$ 
($SCL=1.0$) determines the standard model accretion disk
luminosity which 
follows from $Lum=\sum_{j=1}^{M}{\sigma}T_{\rm eff,j}^4A_j$ where $\sigma$ is the Stefan-Boltzmann 
constant, $T_{\rm eff,j}$ is the effective temperature of the $j^{\rm th}$ annulus, $A_j$ is the radiating area of the
annulus, and there are M annuli. 
Then, having chosen a different value of $EXP$, repeated runs with modified values of $SCL$ eventually identifies
the $SCL$ value producing the same accretion disk luminosity as for $EXP=0.25$.
During each run, application of equation~3 establishes
the $T_{\rm eff}$ values at the various annulus radii assigned by BINSYN. Interpolation among the TLUSTY annulus spectra
(Table~7) produces the BINSYN annulus spectra and integration over them produces the accretion disk synthetic spectrum. 
Since the temperature gradient is smaller for assigned $EXP$ values than the standard model it 
is certain that the TLUSTY 
models will span the
required BINSYN range of models.
We found that the quality of fit improved as
$EXP$ is reduced until we reached the value $EXP=0.125$ but then became poorer as $EXP$ approached 0. Figure~6 shows
the results over the full spectral interval; Figure~7 shows the results in the $FUSE$ region.

To study the quality of fit in detail we have performed
a ${\chi}_{\nu}^2$ analysis with 8 values of $\dot{M}$ (5.0, 5.25, 5.5, 5.75, 6.0, 6.25, 6.5, and 
$7.0{\times}10^{-9}~M_{\odot}~{\rm yr}^{-1}$) and 11 values of $EXP$ (0.25, 0.225, 0.20, 0.175, 0.15, 0.125, 0.10,
0.075, 0.05, 0.025, and 0.0). As explained in connection with the $\sigma_i$ of equation~1, our use of noise level $\sigma_i$
errors would guarantee that the calculated ${\chi}_{\nu}^2$ values will be far larger than 1.0; nevertheless the
calculation preserves relative values enabling us to identify best fit parameters. 
In the overlap region between the $FUSE$ spectrum and the $IUE$ spectrum the $FUSE$ $\sigma_{\rm i}$,
of order $1{\times}10^{-13}~{\rm erg}~{\rm cm}^{-2}~{\rm s}^{-1}~{\rm \AA}^{-1}$, are approximately a
factor 2 smaller than the $IUE$ $\sigma_{\rm i}$. To give equal weight to all the observed points we
used the mean $\sigma_{\rm i}$ of both data sets as a fixed $\sigma_{\rm i}$ in equation~1.
The calculated ${\chi}_{\nu}^2$
values, each with 1086 degrees of freedom, are in Table~9 and a plot is in Figure~8; standard model accretion disks are 
represented by the
horizontal line at the top of the plot.
For ease of calculation we have identified the degrees of freedom with $N$ in equation~1; the effect is a very slight
reduction of ${\chi}_{\nu}^2$ as compared with use of N-M as the degrees of freedom; it is equivalent to masking
an additional 2\AA~in the $FUSE$ spectrum. A separate calculation for $\dot{M}=5.5{\times}10^{-9}M_{\odot}~{\rm yr}^{-1}$
and $EXP=0.125$ and using the individual $FUSE$ and $IUE$ $\sigma_{\rm i}$ values in equation~1 produced a
${\chi}_{\nu}^2$ value of 18.7 in contrast to the value listed in Table~9. Use of the individual $\sigma_{\rm i}$ in
calculating Table~9 would have produced a Figure~8 essentially identical to the one plotted but with slightly larger
contour labels.
We call attention to the fact that the best fit value of $EXP$ is 
close to 0.125 at all mass transfer rates, with a slight downward slope with increasing $\dot{M}$. 
It is curious that the value 0.125 is half of the standard model value, 0.25.

We used a WD $T_{\rm eff}=50,000$K for all models appearing in Table~9. Although the WD contribution to a system
synthetic spectrum is small, it is of interest to verify that the adopted $T_{\rm eff}$ produces optimum
${\chi}_{\nu}^2$ fits. Further, having identified best fit values of $EXP$ across the full range of ${\dot{M}}$
values tested, it is of interest to determine whether
further refinement can identify a preferred value of WD $T_{\rm eff}$. We adopt the value $EXP=0.125$ as fixed and repeat
a ${\chi}_{\nu}^2$ analysis with the same values of $\dot{M}$ as Figure~8 and with 9 values of $T_{\rm eff}$
35,000, 40,000, 45,000, 50,000, 55,000, 60,000, 65,000, 70,000 and 75,000K. 
Since the WD radius is temperature-sensitive we determined $0.90M_{\odot}$ WD radii \citep{panei2000} for 
the various assigned $T_{\rm eff}$
values and used them in calculating the accretion disks. 
The ${\chi}_{\nu}^2$ values are in
Table~10 and
a plot of the results is in Figure~9.
It is clear that the choice of $T_{\rm eff}=50,000$K was about optimum for the available data but that no 
actual determination of the WD $T_{\rm eff}$ is possible. 
   
\section{The orbital inclination, distance, and a best fit model}

The Table~9 and Table~10 results apply for the adopted values of distance (289pc) and $i$ ($34\arcdeg$).

According to BEU the value of $i$ is $34{\arcdeg}{\pm}6{\arcdeg}$. We have tested the effect on the
synthetic spectrum fit, for $\dot{M}=5.75{\times}10^{-9}~M_{\odot}~{\rm yr}^{-1}$, $EXP=0.125$,
by calculating a synthetic model spectrum for the limiting values of $i$ with all other
parameters fixed. For $i=28\arcdeg$, as expected from the angle cosine, the synthetic spectrum was higher by
7\%, and for $i=40\arcdeg$ the synthetic spectrum was lower by 7\%. 
These limiting
spectra differ from the observed spectrum by a small amount, far smaller than the variation from the uncertainty
in the distance, and could be compensated by a small change in the adopted distance without affecting the
quality of fit.	Our results are consistent with $i=34\arcdeg$ but the synthetic spectra cannot constrain $i$.

The~$Hipparcos$~parallax has $1\sigma$ limits of $\pm2.44$mas; the corresponding upper and lower distances are
980pc and 169pc. The RW Sex distance according to BEU is 150pc. 
If the true distance is close to 150pc a smaller synthetic spectrum scaling factor applies and a correspondingly
smaller $\dot{M}$ is required. Experiments with smaller $\dot{M}$ values show that the same standard model spectral
gradient problem persists in those cases as with the simulations discussed above. After several tests the value 
$\dot{M}=2.0{\times}10^{-9}~M_{\odot}~{\rm yr}^{-1}$
appeared the best fit using the 150pc scaling factor. Models with a range of values of $EXP$,
equation~3, led to a visually identified best fit with $EXP=0.15, SCL=0.51443$. It is least laborious to make 
final adjustments in the scaling
factor rather than recalculate models for slightly modified $\dot{M}$; our final scaling factor is $2.5{\times}10^{41}$
corresponding to a distance of 162pc and in agreement with the determination of 150pc.
The corresponding ${\chi}_{\nu}^2$, for comparison with Table~9, was 15.61,
indistinguishable from the optimum value from Table~9.
For comparison, the scaling factor for Figures 4,5,6, and 7 is $9.14{\times}10^{42}$.

We call attention to the fact that the smaller spectral gradient associated with $EXP=0.15$, as compared with
the standard model, avoids the problem of disk instability against ouburst near a $T_{\rm eff}$ of 6000K.
The rim temperature of our best fit model is 7667K while the rim temperature of a corresponding standard
model would be 4216K, well below the limit for onset of outbursts.

Using the K-band 2MASS \citep{skr2006} magnitude of 10.07 and the absolute $M_K$ magnitude for RW Sex 
from \citet{knigge2006,knigge2007}
for an orbital period of $5.88^{\rm h}$, we derive a lower limit distance of 124pc and an upper limit of 217pc,
slightly favoring the distance of 150pc.
\citet{Cowley1977} quote a trigometric parallax determination of 7mas by Osvalds corresponding to a distance
of ${\approx}143$pc. 
The ${\chi}_{\nu}^2$ values do not distinguish between the 289pc and 150pc distances; 
the weight of the evidence favors
150pc which we accept.

Our best fit model has the parameters listed in Table~11.

Figure~10 shows the best fit model together with the various data sources, identified in the figure legend.
Figure~11 shows the same data set for a restricted wavelength range. The observational data plotted are the
values before masking, in contrast to all previous plots. 

Examination of Figure~11 shows that the model fits Lyman $\alpha$ fairly well although the model has a wider
and less deep profile than the HST spectrum. The $FUSE$ Lyman $\beta$ line shows emission peaks in both wings
relative to the model. We suggest this proposed unmodeled effect (emission wings) may be the explanation of those 
residuals and 
that the same explanation
may apply to the Lyman $\gamma$ line. An extension of the argument could explain the remaining shorter wavelength residuals.

Figure~12 is a plot of the residuals from the best fit model to the masked combined spectrum for a restricted wavelength 
range. The plot is a continuous line but the plot data consist of individual values at spacings of approximately 1.5\AA.
Comparison with Figure~10 shows that the synthetic spectrum is much smoother; Doppler shifts in the Keplerian accretion
disk wash out narrow line features in the model spectrum; the residuals in the upper panel of Figure~10 have the appearance 
of noise, with a visually estimated mean of about $2{\times}10^{-13}{\rm erg}~{\rm cm}^{-2}~{\rm s}^{-1}~{\rm \AA}^{-1}$
and frequent larger residuals. The bottom panel of Figure~12 shows that the residuals have a systematic variation
in the 900\AA~to 1200\AA~interval. Excluding that interval temporarily, which includes the entire $FUSE$ spectrum, we test the
hypothesis that the model spectrum is a good fit to the (masked, $IUE$) observed spectrum. 
There are 826 data lines in the spectral interval 
selected; the standard deviation, $\sigma_{\rm r}$, of the residuals is 
$2.22{\times}10^{-13}~{\rm erg}~{\rm cm}^{-2}~{\rm s}^{-1}~{\rm \AA}^{-1}$, the value of ${\chi}^2$ is 828.62 and the value of
${\chi}_{\nu}^2$ is 1.009 for 821 degrees of freedom. Using the incomplete gamma function routine GAMMQ 
from \citet{press} we calculate the
probability Q that the ${\chi}^2$ value as poor as 828.62 could occur by chance is 0.419; according to \citet{press}
a value Q greater than 0.1 indicates an acceptable model. We also apply the Kolmogorov-Smirnov test where the null
hypotheis is that the residuals represent a normal distribution with mean of zero. The test determined the $\sigma_{\rm r}$
listed above and sorted the residuals into 60 bins, 58 of width $0.1\sigma_{\rm r}$, one containing all residuals greater
than $2.55\sigma_{\rm r}$ and the other containing all residuals smaller than $-2.55\sigma_{\rm r}$. The program calculated a 
theoretical cumulative distribution function, the actual cumulative distribution function of the residuals, and
the difference between the two. The maximum difference was 0.168 while the critical value 
at the 0.05 level of significance \citep[Table~16]{ostle}
was 0.0473; the hypothesis (normal distribution with mean of zero) is rejected.

The maximum systematic residuals of the $FUSE$ spectrum, Figure~12, bottom panel, are of order 1/3 of the observed spectrum
flux values. The best fit model, essentially identical with the magenta line of Figure~7, falls below the
$FUSE$ spectrum shortward of 1050\AA~and has greater flux longward.	We have already speculated that there may be
unmodeled emission in the shorter wavelengths. We found (\S4.) that the $FUSE$ spectrum apparently
was anomalously weak at the time of observation	and a correction factor was necessary to combine it with the $IUE$
spectrum.

\subsection{How robust is the best fit model?}

The model parameters subject to adjustment are $M_{\rm wd},~T_{\rm eff,wd},~i,~\dot{M},~\rm{distance},~{\rm and}~EXP$.
As explained above, variation of $i$ produces a minor effect at the adopted $i$ and in any case $i$ cannot be used to
constrain the model because RW Sex is a non-eclipsing system. 

Variation of $M_{\rm wd}$ has two effects; first, there is
a variation of WD radius with mass, leading to variation of the WD luminous flux for a given WD $T_{\rm eff}$, and second,
the variation of the depth of the potential well produces variation of annulus $T_{\rm eff}$ at a given distance from the
WD for a given $\dot{M}$. As shown by, e.g., Figure~10, the WD contribution to the system synthetic spectrum is small
and the variation of it (the synthetic spectrum) due to a change in $M_{\rm wd}$ is second order; we neglect it in comparison with other major
contributors to the model spectrum. Variation of an annulus $T_{\rm eff}$ from variation of $M_{\rm wd}$ can be included
in the study of $\dot{M}$ effects since the latter also produce variation of annulus $T_{\rm eff}$ at a given distance
from the WD and the variation with $\dot{M}$ is the dominant effect. 
We exclude $M_{\rm wd}$ as a significant contributing source to the discrepancy between observation and
the standard model.

If we reset $EXP=0.25$ (i.e., assert the standard model is applicable), can we vary the RW Sex distance or $\dot{M}$ or the
WD $T_{\rm eff}$ and achieve as good a fit to the observed spectra as our model does in Figure~10 and Figure~11?
From Figure~4, the closest synthetic spectrum fit
to the observations is with the smallest $\dot{M}$, $5.0{\times}10^{-9}M_{\odot}~{\rm yr}^{-1}$. Still smaller $\dot{M}$
values, tested through $1.0{\times}10^{-9}M_{\odot}~{\rm yr}^{-1}$,
show the same trend and ultimately the $\dot{M}$ value becomes small enough that the accretion disk becomes
unstable against outburst. At $\dot{M}=2.0{\times}10^{-9}M_{\odot}~{\rm yr}^{-1}$ the outer accretion disk rim
has a calculated $T_{\rm eff}$ of 4215K, making it susceptable to outbursts. 
However, stream impact and tidal effects heat the outer disk region \citep{lasota2001}; let us assume these effects stabilize
the standard model disk (and ignore the consequent departure from the standard model)
and inquire how well the standard model for $\dot{M}=2.0{\times}10^{-9}M_{\odot}~{\rm yr}^{-1}$
fits the observations.
Figure~13 shows the fit of the standard model to the observations. The system synthetic spectrum,
including the contributions of both the accretion disk and the WD, 
has been divided by
$3.0{\times}10^{41}$, corresponding to a distance of 180pc.	A larger divisor shifts the synthetic spectrum downward,
producing larger long wavelength discrepancies, and a smaller divisor shifts the synthetic spectrum upward, producing
larger short wavelength discrepancies. The contribution of the 50,000K WD is at the bottom. Removing its contribution
entirely would only marginally reduce the discrepancy from the system synthetic spectrum. A proposed higher $T_{\rm eff}$
WD would raise the question of its source of heat but, in any case, the rapidly increasing WD FUV flux would only make
the system synthetic spectrum discrepancy worse. Still smaller $\dot{M}$ values exhibit the same discrepancy 
shown in Figure~13 and, in any case,
are not credible because of the instability against outburst (the standard model discrepancy from the 
roughly 6000K crossover
to instability becomes worse and the proposed stream impact heating becomes smaller with reduced $\dot{M}$).
Figure~13 contrasts strongly with the very good fit of our model in Figure~10. 
We argue that
no standard model synthetic spectrum, for any acceptable $\dot{M}$, fits the observed spectrum and that
variation of the RW Sex distance, accompanied by a compensating variation in $\dot{M}$
will not produce an acceptable standard model fit to the observed spectra. 

We also argue, from the above discussion of Figure~13, that
variation of the WD $T_{\rm eff}$ cannot be used to produce an acceptable standard
model fit to the observed spectrum. 
Addition of a boundary layer with a long wavelength tail in the FUV would only make the Figure~13 fit worse.
In Figure~11 the synthetic spectrum falls below the observations near the FUV limit and addition of a
boundary layer contribution could improve the fit. However we have already suggested that emission from the
chromosphere could explain the discrepancy and we believe the evidence is too weak to support a proposal
for a boundary layer.

We are left with variation of $EXP$, from our previous discussion, as the only 
viable parameter to produce
an acceptable fit to the observed RW Sex spectrum.   

\section{The time-series $FUSE$ spectra}

\citet{prinja2003} show that there is large orbit-to-orbit variability of the time-series $FUSE$ spectra. 
The changes are primarily in the absorption strengths of the high excitation lines, associated with variable
blue shifts tied to the orbital period.
\citet{prinja2003} argue that the variation arises from blueshifted absorptive changes as opposed to a
blueshifted emission component and they suggest that the outflow wind is oblique rather than a symmetric bipolar wind
with the oblique wind possibly seated on a warped or tilted disk.

We have used our best fit model to test the tilted disk suggestion. We first subtracted our model synthetic spectrum, 
scaled to the distance
of RW Sex, from the 25 individual time-series spectra. 
This step produces residuals which would display phase-wise flux variations if the disk is tilted.
We then chose one difference spectrum (the second in the sequence) 
and successively overplotted the other difference spectra. This comparison verified the large case-to-case variation
in the strengths of the absorption lines as well as the variable blue shift but there was no indication of a vertical
shift in successively overplotted difference spectra. If the disk is warped or tilted there should be a vertical displacement
of the entire difference spectrum that is tied to the orbital phase and we find none. 
Figure~14 and Figure~15 illustrate the comparison and have the most striking differences. Superposition of the black 
difference spectrum and other
difference spectra of similar orbital phase shows close matches, indicating that the spectral changes primarily
connect to orbital phase. 
  
This result agrees with
the \citet{prinja2003} proposal that the spectral variation arises from blueshifted absorptive changes.
As an alternative to the the tilted disk suggestion, we believe a likely explanation of the non-symmetrical wind is an
interaction with the debris from the mass transfer stream impact on the accretion disk.
\citet{lubow} find that the stream impact on the accretion disk leads to material flowing over the
accretion disk from the impact location.
 Hydrodynamical 2D models of the stream
 impact \citep{rozy1987,rozy1988} identify two shock waves: (1) a shock on a plane perpendicular to the
 orbital plane, roughly bisecting the
 angle between the stream and the rim and terminating at the upstream edge of the stream, and (2) a 
 shock slightly more inclined to the stream and extending far into the disk. Although the simulation is
 2D, \citet{rozy1987} states that a bow shock will develop, prospectively leading to vertical expansion
 upstream. \citet{livio1986} and \citet{ar1998} perform a 3D simulation and find that material from the 
 stream flows over
 the disk if cooling is efficient, applicable to low $\dot{M}$ cases, and is more like an explosion in
 high $\dot{M}$ cases, leading to a bulge extending along the disk rim.
 
 Impact of the mass transfer stream on the accretion disk might be expected to produce a detectable rim hot spot, but
 the absence of phase-wise flux variability demonstrates that evidence is lacking for a 
 sufficiently luminous rim hot
 spot to be detectable. This justifies our omission of a hot spot in the system model (\S6).

\section{Discussion}

As discusssed in the Introduction, \citet{wade1988} showed that neither steady state model accretion disks based
on Planck functions nor stellar model atmospheres could simultaneously fit the colors and absolute
luminosities of a set of NL systems (RW Sex was included in the study). Tomographic analyses of 
accretion disks \citep{rut1992}
show clear departures from standard model temperature profiles (FKR), with a spectral gradient typically
less steep than for a standard model.
Recent analyses of the NL systems MV Lyr, IX Vel, QU Car, and UX UMa
\citep{linnell2005,linnell2007,linnell2008a,linnell2008b} used an annulus model explicitly representing
standard model accretion disks and
showed, in each case,
that a standard model accretion disk synthetic spectrum could not accurately fit observed spectra.
On the other hand, a standard model
fits the observed spectrum in V3885 Sgr \citep{linnell2009}.
In the latter paper we speculated that a given system may at times fit a standard model and at other times
show departures as observed in the case of UX UMa $IUE$ spectra.

In the specific cases of non-standard model accretion disks listed above, the fits achieved depended on ad hoc changes
in the temperatures of a few accretion disk annuli.	The required departure from a standard model accretion disk
temperature profile to fit a model synthetic spectrum to an observed spectrum has not previously been characterized 
analytically.
This study shows, by detailed ${\chi}_{\nu}^2$ analysis, that the observed RW Sex spectrum accurately
matches a synthetic spectrum which follows from
an analytic expression, equation~3, that differs from the standard model.
Referring back to the \citet{wade1988} results, we conclude the problem is not a failure of synthetic spectra to represent
the standard model but rather observed spectra, here RX Sex, fail to conform to the standard model.
A further consequence is that the lower temperature gradient produces a higher rim $T_{\rm eff}$, preserving
stability against outburst to a lower $\dot{M}$.
  
We have taken extensive precautions to consider physical effects that might otherwise affect the accuracy
of the calculated synthetic spectra: the annulus synthetic spectra are non-LTE and important metals
have been included as explicit ions in the calculations.
It is curious that the empirically-determined exponent in equation~3, equal to 0.25 for the standard model, is
closely equal to 1/2 of the standard model value for a range of $\dot{M}$ values. 
Equation~3 was not derived from a physical analysis and it is of interest to consider its implications.
The less steep temperature profile than the standard model suggests additional energy deposition terms
in a given annulus than considered in the standard model, such as radial energy transfer between annuli
\citep{p1995}, but it seems intuitively doubtful that the latter effect could
produce the large departure observed. In any case, that physical effect would be present for all
accretion disks and would leave the case-to-case differences unexplained.
Our assumption of a constant $\alpha$ for the
entire accretion disk could be challenged but, as FKR discuss in connection with their equation 5.18,
the energy flux through an accretion disk face, in the standard model, is independent of viscosity
and is not an explicit function of $\alpha$: to first order the radial temperature profile is independent
of $\alpha$. 
The derivation of equation~2 neglects possible magnetic effects but their inclusion, as mentioned in \S6,
currently is subject to some uncertainty.  
Further consideration of this topic is beyond the scope of this paper.

We have neglected irradiation of the annuli by the WD. \citet{smak1989} shows that the effects are small
for stationary disks. In RW Sex the presence of a disk chromosphere would reduce irradiation effects
still further.  

It is of interest to place the results of this paper in the context of other approaches to the departure
of observational data from the standard model. IX Vel provides an illustration; \citet{long1994}
consider several options including removal of all radiative flux contribution from inner annuli, providing
a constant $T_{\rm eff}$ inner region, and setting the formal inner radius of the accretion disk to
some multiple of the WD radius. These authors show that the last option, with $R_{\rm min} {\sim} 2.6R_{\rm wd}$,
gives a reasonable fit to the HUT (Hopkins Ultraviolet Telescope) data. \citet{linnell2007} showed that
(for IX Vel)
an accretion disk with a constant temperature inner section and the remainder following the standard model
gives a good fit to $FUSE$ and STIS spectra. Both approaches are ad hoc and the truncated disk model
leaves unanswered the question of whether a wind can carry away all of the mass transfer stream,
starting at the truncation radius. The ad hoc processes are localized perturbations of an otherwise
standard model accretion disk; the analytic representation found here suggests existence of a process that affects
the entire accretion disk. 

Our model does not include a boundary layer (BL).
As discussed by FKR, half of the potential energy liberated in the fall of the mass transfer stream
from the L1 point (essentially from infinity) to the WD surface appears as radiated energy. Energy 
conservation requires an accounting for the other half and the presence of a hot BL is
a common prescription. A problem is that BLs with the prescribed properties typically are not
observed in high $\dot{M}$ systems or occur at unobserved wavelengths \citep{cordova1995}. 
The absence of an observed but predicted (FKR) BL in high $\dot{M}$ CV systems has an 
extensive history: 
\citep{ferland1982,kaljen1985,
patray1985,hdr1991,hdr1993,vrt1994,idan1996}.

In the case of the low $\dot{M}$ dwarf nova (DN) U Gem, \citet{long1996} used eclipse data during outburst 
to determine that an
emitting region is present with a temperature of ${\sim}$140,000K, and a size approximating that
of the WD, consistent with a boundary
layer with a luminosity comparable to the disk luminosity. \citet{szkody1996} also found that in quiescence 
U Gem has a relatively hard X-ray spectrum with the emission confined to a small area, supporting
the interpretation of a boundary layer.
More recently, \citet{pandel2005} show that, for 9 DN in quiescence,
X-ray observations require a model in which a hot boundary layer is present and has a luminosity approximately
equal to the accretion disk luminosity. In the cases cited by \citet{pandel2005} there is no boundary layer problem and
the derived boundary layer structure is consistent with theoretical models \citep{narayan1993}.
The derived $\dot{M}$ rates for the DN in quiescence are 
$\dot{M} {\sim} 10^{-12}~{\rm to}~10^{-11}M_{\odot}~{\rm yr}^{-1}$.
The $\dot{M}$ rate for RW Sex is between two and three dex larger.
  
One proposed explanation for lack of evidence for a high luminosity boundary layer is that the WD is 
rotating close to Keplerian rotation, with no
BL predicted. However, HST measurements
starting with \citet{Sion1994} have measured rotational WD velocities too low to explain "missing"
BLs.
An alternative explanatory theme is that the absence of an
observed BL
associates with a wind, and in RW Sex a wind is
clearly present \citep{prinja2003}, but the adequacy of this explanation remains undemonstrated.

\section{Summary}

The BINSYN program suite has been used to calculate model synthetic spectra for RW Sex in comparison with a
combined HST, $FUSE$ and $IUE$ spectrum. 
From evidence reported in this paper,
the $FUSE$ spectrum was obtained, apparently, at a time of reduced accretion disk luminosity and 
to achieve consistency it is necessary
to divide the $FUSE$ spectrum by 0.82 in combining it with the other observed spectra.

The models include a range of $\dot{M}$ values and each model
consists of 45 annuli calculated with the Hubeny program TLUSTY; the Hubeny program SYNSPEC produces 
the synthetic spectrum for a
single annulus, with a resolution of 0.1\AA, includes
significant metals as explicit ions, and is a non-LTE model for all annuli that contribute significantly
to the system synthetic spectrum.
A~$Hipparcos$~parallax, determining $D=289$pc, fixes the divisor that enables the system synthetic spectrum to be
matched to the observed spectrum and establishes the fit on an absolute flux basis.
Initial tests with a range of $\dot{M}$ standard model accretion disks show that no
standard model can fit the observed spectra of RW Sex. 
New models with a generalized form of the standard model equation
show that, for RW Sex, and based on a ${\chi}_{\nu}^2$ analysis,
the exponent which determines the radial temperature profile is approximately 0.125
rather than the standard value 0.25.
The optimum $\dot{M}$ for $D=289$pc is $\dot{M}=5.75{\times}10^{-9}~M_{\odot}~{\rm yr}^{-1}$, assuming
the WD $T_{\rm eff}=50,000$K. A separate ${\chi}_{\nu}^2$ analysis shows that the data sensitivity
to the WD $T_{\rm eff}$ is too small to permit an actual determination of the WD $T_{\rm eff}$ but that a value
of 50,000K is a reasonable choice.

Tests based on a separately determined $D=150$pc, more recent than the $Hipparcos$ determination, show 
that a non-standard model with
$\dot{M}=2.0{\times}10^{-9}~M_{\odot}~{\rm yr}^{-1}$ produces a comparable synthetic spectrum
fit to the observed spectrum as for $D=289$pc ($Hipparcos$) and the ${\chi}_{\nu}^2$ difference 
in the two cases is too small to be significant. We tabulate a best fit model for the $D=150$pc case.

The residual spectra between individual time-series spectra (whose average is the $FUSE$ spectrum described
above) and the best fit synthetic spectrum model permit
a study of the proposal for a warped or tilted accretion disk. The absence of a phase-synchronized vertical shift
in the difference spectra rules out a tilted disk. An alternative proposal for the asymmetric wind is an 
interaction between the wind and the debris from the mass transfer stream impact on the accretion disk. 

We are grateful for the careful reading of the manuscript by the referee and his/her insightful comments
and criticisms, leading to an improved presentation. 

P.S. acknowledges support from NSF grant AST-0607840. 
Support for this work was provided by NASA through grant number 
HST-AR-10657.01-A  to Villanova University (P. Godon) from the Space
Telescope Science Institute, which is operated by the Association of
Universities for Research in Astronomy, Incorporated, under NASA
contact NAS5-26555. Additional support for this work was provided by the National Aeronautics
and Space Administration (NASA) under Grant number NNX08AJ39G issued through the office
of Astrophysics Data Analysis Program (ADP) to Villanova University (P. Godon).
Participation by E.~M.~Sion, P.~Godon and A. Linnell was also supported in part by NSF grant AST0807892
to Villanova University.
This research was partly based on observations made with
the NASA-CNES-CSA Far Ultraviolet Spectroscopic Explorer. $FUSE$ is operated
for NASA by the Johns Hopkins University under NASA contract NAS5-32958.

\clearpage



\clearpage


\begin{deluxetable}{lccrcccc}
\tabletypesize{\small}
\tablewidth{0pc}
\tablecaption{FUV Observations of RW Sex: FUSE \& HST Spectra}
\tablehead{
Instrument  & Date      &  time      & Exp.time   & Dataset   & Aperture & Operation   & Wavelengths   \\ 
            &(dd/mm/yy) & (hh:mm:ss) & (sec)$~~~$ &           & or Grating & Mode      &  (\AA )    
}
\startdata 
FUSE       & 13-05-01   & 14:48:05   & 25,614     & B1040101  & LWRS     & TTAG       &  904-1188   \\
HST/GHRS   & 04-05-96   & 13:03:53   &   544      & z37v0104t & G140L    & ACCUM      & 1367-1663   \\ 
HST/GHRS   & 04-05-96   & 13:17:58   &   544      & z37v0105t & G140L    & ACCUM      & 1367-1663   \\ 
HST/GHRS   & 04-05-96   & 13:30:59   &   544      & z37v0106t & G140L    & ACCUM      & 1140-1435   \\ 
HST/GHRS   & 04-05-96   & 13:37:23   &   544      & z37v0107t & G140L    & ACCUM      & 1140-1435    \\ 
HST/GHRS   & 04-05-96   & 14:50:24   &   544      & z37v0108t & G140L    & ACCUM      & 1367-1663   \\ 
HST/GHRS   & 04-05-96   & 15:03:25   &   435      & z37v0109t & G140L    & ACCUM      & 1367-1663   \\ 
\enddata 
\end{deluxetable}


\clearpage

 
\begin{deluxetable}{lccc}
\tablecaption{{\it{FUSE}} ISM Line Identifications}
\tablehead{ 
Ion          & $\lambda_{\rm rest}$(\AA)&$\lambda_{\rm obs}$(\AA) & Comments    
}
\startdata 
H\,{\sc i}   & 914.29    &  914.35   &  18u   \\  
H\,{\sc i}   & 914.57    &  914.58   &  17u   \\  
H\,{\sc i}   & 914.92    &  914.95   &  16u   \\  
H\,{\sc i}   & 915.33    &  915.34   &  15u   \\  
H\,{\sc i}   & 915.82    &  915.85   &  14u   \\  
H\,{\sc i}   & 916.43    &  916.47   &  13u   \\  
O\,{\sc i}   & 916.82    &  916.90   &  26u   \\  
H\,{\sc i}   & 917.18    &  917.22   &  12u   \\  
H\,{\sc i}   & 918.13    &  918.18   &  11u   \\  
O\,{\sc i}   & 918.29    &  918.48   &  25u  \\  
O\,{\sc i}   & 918.87    &  918.90   &  25u   \\  
H\,{\sc i}   & 919.35    &  919.42   &  10u   \\  
...          &           &           &        \\ 
O\,{\sc i}   & 948.69    &  948.70   &   12u   \\  
H\,{\sc i}   & 949.74    &  949.80   &   4u   \\  
O\,{\sc i}   & 950.89    &  950.90   &   11u   \\  
P\,{\sc ii}  & 961.04    &  961.10   &         \\  
P\,{\sc ii}  & 962.12    &  962.20   &         \\  
P\,{\sc ii}  & 962.57    &  962.65   &         \\  
P\,{\sc ii}  & 963.62    &  963.65   &         \\  
P\,{\sc ii}  & 963.80    &  963.85   &         \\  
N\,{\sc i}   & 963.99    &  964.05   &   3u    \\  
Fe\,{\sc ii} & 964.30    &  964.37   &         \\  
N\,{\sc i}   & 964.63    &  964.70   &   3u    \\  
P\,{\sc ii}  & 964.95    &  965.07   &         \\  
N\,{\sc i}   & 965.05    &  965.07   &   3u    \\  
Fe\,{\sc ii} & 966.20    &  966.27   &         \\  
O\,{\sc i}   & 971.74    &  971.80   &   10u   \\  
H\,{\sc i}   & 972.54    &  972.65   &   3u   \\  
O\,{\sc i}   & 976.45    &  976.50   &   7u   \\  
C\,{\sc iii} &    977.02 &  977.10   & 1u sharp absorption     \\ 
O\,{\sc i}   & 988.66    &  988.70   &   5u   \\  
N\,{\sc iii} & 989.80    &  989.90   &   1u    \\  
Si\,{\sc ii} & 989.87    &  989.90   &   6u    \\  
S\,{\sc i}   & 997.01    &  997.15   &         \\  
O\,{\sc i}   & 1025.76   & 1025.90   &   4u   \\  
C\,{\sc ii}  &   1036.34 & 1036.35   &   2u      \\
C\,{\sc ii}  &   1037.02 & 1037.15   &   2u      \\
O\,{\sc i}   & 1039.23   & 1039.30   &   3u   \\  
Ar\,{\sc i}  &   1048.20 & 1048.25   &   2u     \\
Ar\,{\sc i}  &   1066.66 & 1066.70   &   1u      \\
N\,{\sc ii}  & 1083.99 & 1084.05   &  1u  \\
N\,{\sc ii}  & 1084.57 & 1084.60   &  1u  \\
Fe\,{\sc ii} & 1096.61 & 1096.65   &  18u \\
Fe\,{\sc ii} & 1096.78 & 1096.80   &  18u \\
Fe\,{\sc ii} & 1112.47 & 1112.48   &      \\
Fe\,{\sc ii} & 1121.97 & 1121.9    &      \\
Fe\,{\sc ii} & 1125.45 & 1125.4    &      \\
N\,{\sc i}   &  1134.17 & 1134.10   & 2u   \\
N\,{\sc i}   &  1134.42 & 1134.37   & 2u   \\ 
N\,{\sc i}   &  1134.98 & 1134.92   & 2u     \\ 
Fe\,{\sc ii} & 1143.23 & 1143.18   &  10u \\
Fe\,{\sc ii} & 1144.94 & 1144.88   &  10u \\
P\,{\sc ii}  & 1152.82 & 1152.75   &  3u  \\
?            & 1181.35? & 1181.40   &                   \\  
\enddata 

\tablecomments
{In the comments column, letter u stands for 
an ultraviolet multiplet and the preceeding number is the
multiplet number. See \citet{morton2000,morton2003}.
}   
\end{deluxetable}


\clearpage

 
\begin{deluxetable}{lccc}
\tablecaption{{\it{FUSE}} Line Identifications: The blue-shifted broad
absorption lines}
\tablehead{ 
Ion          & $\lambda_{\rm rest}$(\AA)&$\lambda_{\rm obs}$(\AA) co-added & 
$\lambda_{\rm obs}$(\AA) diff.      
}
\startdata 
N\,{\sc iv}  &    921.46 & $\sim$920.0 & $\sim 921.5$     \\
             &    921.99 & $\sim$920.0 & $\sim 921.5$     \\
             &    922.52 & $\sim$920.0 & $\sim 921.5$     \\
             &    923.06 & $\sim$920.0 & $\sim 921.5$     \\
             &    923.22 & $\sim$920.0 & $\sim 921.5$     \\
             &    924.28 & $\sim$920.0 & $\sim 921.5$     \\
             &    924.91 & $\sim$920.0 & $\sim 921.5$     \\
S\,{\sc vi}  &    933.38 & $\sim$930.  & $\sim 932.0$     \\
             &    944.52 & $\sim$940.9 & $\sim 942.3$     \\
P\,{\sc iv}  &    950.66 & $\sim$947.0 & $\sim 949.2$     \\ 
C\,{\sc iii} &    977.02 & $\sim$973.5 & $\sim 975.2$     \\ 
N\,{\sc iii} & 989.80    & $\sim$988.0 & $\sim 988.4$     \\  
N\,{\sc iii} & 991.51    & $\sim$988.0 & $\sim 990.4$     \\  
N\,{\sc iii} & 991.58    & $\sim$988.0 & $\sim 990.4$     \\  
O\,{\sc vi}  &   1031.91 & 1028.9?     & $\sim 1029.6$     \\
             &   1037.61 & 1034.6?     & $\sim 1035.9$     \\
S\,{\sc iv}  & 1062.65 & $\sim$1060.7  & $\sim 1061.4$     \\ 
S\,{\sc iv}  & 1072.97 & $\sim$1070.4  & $\sim 1071.8$     \\ 
             & 1073.52 & $\sim$1070.4  & $\sim 1071.8$     \\ 
P\,{\sc v}   & 1117.98 & $\sim$1115.2  & $\sim 1116.5$     \\ 
P\,{\sc iv}  & 1118.55 & $\sim$1115.2  & $\sim 1116.5$     \\ 
Si\,{\sc iv} & 1122.49 & $\sim$1120.0  &  --- \\ 
P\,{\sc v}   & 1128.01 & $\sim$1125.4  & $\sim 1126.6$     \\ 
Si\,{\sc iv} & 1128.33 & $\sim$1125.4  &  --- \\ 
Si\,{\sc iii} & 1144.31  & 1142.7      & --   \\ 
              & 1144.96  & 1142.7      & ---  \\ 
              & 1145.11  & 1142.7      & ---  \\ 
              & 1145.18  & 1142.7      & ---  \\ 
              & 1145.67  & 1142.7      & ---  \\ 
C\,{\sc iii} &  1174.90 & 1172.7       & $\sim 1174.1$     \\ 
             &  1175.26 & 1172.7       & $\sim 1174.1$     \\ 
             &  1175.60 & 1172.7       & $\sim 1174.1$     \\ 
             &  1175.71 & 1172.7       & $\sim 1174.1$     \\ 
             &  1176.00 & 1172.7       & $\sim 1174.1$     \\ 
             &  1176.40 & 1172.7       & $\sim 1174.1$ 
\enddata    
\end{deluxetable}



 
\begin{deluxetable}{lcccl}
\tablecaption{{HST/\it{GHRS}} Line Identifications}
\tablehead{ 
Ion          & $\lambda_{\rm rest}$(\AA)&$\lambda_{\rm obs}$(\AA) & $\sim$shift (\AA) & Origin
}
\startdata 
Si\,{\sc iii} & 1144.31 & $\sim 1142.5$ & -2.5 & source          \\ 
              & 1144.96 & $\sim 1142.5$ & -2.5 &      \\ 
              & 1145.11 & $\sim 1142.5$ & -2.5 &      \\ 
              & 1145.18 & $\sim 1142.5$ & -2.5 &      \\ 
              & 1145.67 & $\sim 1142.5$ & -2.5 &      \\ 
P\,{\sc ii}   & 1142.89 &       1142.5  & -0.4 & ISM              \\
P\,{\sc ii}   & 1152.82 &       1152.6  & -0.2 & ISM              \\
C\,{\sc iii}  & 1174.90 & $\sim 1173.4$ & -2.5 & source    \\ 
              & 1175.26 & $\sim 1173.4$ & -2.5 &   \\ 
              & 1175.60 & $\sim 1173.4$ & -2.5 &   \\ 
              & 1175.71 & $\sim 1173.4$ & -2.5 &   \\ 
              & 1176.00 & $\sim 1173.4$ & -2.5 &   \\ 
              & 1176.40 & $\sim 1173.4$ & -2.5 &  \\  
Si\,{\sc ii}  & 1190.42 & 1190.1        & -0.3 &  ISM              \\ 
Si\,{\sc ii}  & 1193.29 & 1192.9        & -0.4 &  ISM             \\  
N\,{\sc i}    & 1199.55 & $\sim 1200.0$ & -0.2 & ISM, unresolved   \\
              & 1200.22 & $\sim 1200.0$ & -0.2 &                  \\
              & 1200.71 & $\sim 1200.0$ & -0.2 &                  \\
Si\,{\sc iii} & 1206.51 & $\sim 1205.0$ & -2.0 & source                    \\
              & 1207.52 & $\sim 1205.0$ & -2.0 &                  \\
H\,{\sc i}    & 1215.67 & $\sim 1215.2$ & -0.5 & source         \\ 
N\,{\sc v}    & 1238.82 & $\sim 1234.2$ & -4.6 & source         \\
              & 1242.80 & $\sim 1237.9$ & -4.9 &                 \\  
S\,{\sc ii}   & 1250.58 & 1250.2        & -0.4 & ISM              \\  
S\,{\sc ii}   & 1253.81 & 1253.5        & -0.3 & ISM              \\  
S\,{\sc ii}   & 1259.52 & 1259.1        & -0.4 & ISM              \\  
Si\,{\sc ii}  & 1260.42 & 1260.0        & -0.4 & ISM               \\
S\,{\sc i}    & 1277.22 & 1277.0        & -0.2 & ISM               \\ 
C\,{\sc i}    & 1277.26 & 1277.0        & -0.3 & ISM               \\ 
Si\,{\sc iii} & 1294.55 & $\sim 1300.0$ & -2.0 & source            \\ 
              & 1296.73 & $\sim 1300.0$ & -2.0 &                  \\  
              & 1298.89 & $\sim 1300.0$ & -2.0 &                  \\  
              & 1298.95 & $\sim 1300.0$ & -2.0 &                  \\  
              & 1301.15 & $\sim 1300.0$ & -2.0 &                  \\  
              & 1303.32 & $\sim 1300.0$ & -2.0 &                  \\  
P\,{\sc ii}   & 1301.87 & 1301.7        & -0.2 & ISM               \\ 
O\,{\sc i}    & 1302.17 & 1301.7        & -0.5 & ISM               \\ 
Si\,{\sc ii}  & 1304.37 & 1303.9        & -0.5 & ISM             \\ 
Si\,{\sc iii} & 1328.81 & $\sim 1327.0$ & -1.8 & source    \\ 
C\,{\sc ii}   & 1334.53 & 1334.2        & -0.3 & ISM           \\ 
C\,{\sc ii}   & 1335.70 & 1335.3        & -0.4 & ISM                \\ 
Si\,{\sc iv}  & 1393.76 & $\sim 1392.0$ & -1.8 & source                 \\ 
              & 1402.77 & $\sim 1401.0$ & -1.8 & source                 \\ 
Si\,{\sc ii}  & 1526.71 & 1526.5        & -0.2 & ISM             \\ 
C\,{\sc iv}   & 1548.19 & $\sim 1542.2$ & -6.0 & source            \\ 
              & 1550.77 & $\sim 1544.7$ & -6.1 & source            \\ 
He\,{\sc ii}  & 1640.40 & $\sim 1638.6$ & -1.8 & source             
\enddata    
\small{             
All the lines from the source 
are broad, asymmetric, blue-shifted by 2 to 6\AA\ , and belong to
higher order ionization species, indicating
that they form in a hot expanding corona above the disk
and/or white dwarf. The lines from the ISM are sharper, blue-shifted by 
0.2-0.5\AA\ , and belong to lower order ionization species.
}    
\end{deluxetable}

\clearpage

\begin{deluxetable}{ccrcccc}
\tabletypesize{\small}
\tablewidth{0pc}
\tablecaption{IUE Observations of RW Sex}
\tablehead{
Date      &  time      & Exp.time   & Description   & Aperture    & Wavelengths   \\ 
(dd/mm/yy) & (hh:mm:ss) & (sec)$~~$    &               &             &  (\AA )    	  \\
}
\startdata 
30-05-78   & 16:45:00   &  1230      & LWR01583  & LARGE    & 1851-3350   \\ 
30-11-78   & 07:06:07   &   714      & LWR03071  & LARGE    & 1851-3350   \\ 
29-12-79   & 19:44:20   &   900      & LWR06494  & LARGE    & 1851-3350   \\
30-05-78   & 18:39:00   &   360      & SWP01671  & LARGE    & 1150-1979   \\ 
30-11-78   & 08:07:12   &   714      & SWP03494  & LARGE    & 1150-1979   \\ 
29-12-79   & 19:00:42   &   900      & SWP07500  & LARGE    & 1150-1979    \\ 
\enddata 
\end{deluxetable}

\clearpage
\begin{deluxetable}{llll}
\tablewidth{0pt}
\tablenum{6}
\tablecaption{RW Sex Initial System Parameters}
\tablehead{
\colhead{parameter} & \colhead{value} & \colhead{parameter} & \colhead{value}}
\startdata
$M_{\rm wd}$  &  $0.90M_{\odot}$\tablenotemark{a}	 & $i$    &  $34{\arcdeg}$~\tablenotemark{d}\\
${M}_2$  &  $0.674{M}_{\odot}$\tablenotemark{b}	 & ${\dot{M}}$ 	&$1.0{\times}10^{-9}{M}_{\odot} 
{\rm yr}^{-1}$~\tablenotemark{e} \\
P    &  0.24507 day~\tablenotemark{c}	  & $d$     	& $289$~pc~\tablenotemark{f}\\
\enddata
\tablenotetext{a}{\citet{beuermann} value of $1/q=1.35$ and ${M}_2$	from
\citet{knigge2006,knigge2007}}
\tablenotetext{b}{\citet{knigge2006,knigge2007} calibrated $P:M_2$ relation}
\tablenotetext{c}{\citet{beuermann}}
\tablenotetext{d}{\citet{beuermann}, $i$ between $28{\arcdeg}$ and $40{\arcdeg}$}
\tablenotetext{e}{arbitrary initial test value}
\tablenotetext{f}{Hipparcos value}
\end{deluxetable}




\begin{deluxetable}{rrrrrrrr}
\tablewidth{0pt}
\tablenum{7}
\tablecaption{Properties of accretion disk with mass transfer rate 
$\dot{M}=5.0{\times}10^{-9}~{M}_{\odot}{\rm yr}^{-1}$ and WD mass of $0.90{M}_{\odot}$.}
\tablehead{	  
\colhead{$r/r_{\rm wd,0}$} & \colhead{$T_{\rm eff}$} & \colhead{$m_0$} 
& \colhead{log~$g$}
& \colhead{$z_0$} & \colhead{$Ne$} & \colhead{{$\tau_{\rm Ross}$}}
}
\startdata
1.36  &  66422  &  1.21E4    &  7.07   & 5.65E7  & 2.87E17  & 1.76E4\\
2.00  &	 59554  &  1.48E4	 &  6.80   & 9.79E7  & 1.89E17  & 2.09E4\\
3.00  &	 48157  &  1.44E4	 &  6.50   & 1.66E8  & 1.11E17  & 2.47E4\\
4.00  &	 40477  &  1.35E4	 &  6.29   & 2.38E8  & 7.39E16  & 2.81E4\\
5.00  &	 35109  &  1.25E4	 &  6.12   & 3.14E8  & 5.35E16  & 3.16E4\\
6.00  &	 31148  &  1.17E4	 &  5.98   & 3.93E8  & 4.11E16  & 3.53E4\\
7.00  &  28096  &  1.10E4    &  5.86   & 4.75E8  & 3.32E16  & 3.93E4\\
8.00  &	 25664  &  1.04E4	 &  5.75   & 5.60E8  & 2.79E16  & 4.34E4\\
9.00  &  23676  &  9.83E3    &  5.66   & 6.42E8  & 2.92E16  & 4.78E4\\
10.00 &	 22016  &  9.37E3	 &  5.58   & 7.28E8  & 2.55E16  & 5.23E4\\
12.00 &	 19393  &  8.60E3	 &  5.44   & 9.08E8  & 2.03E16  & 6.18E4\\
14.00 &	 17404  &  7.98E3	 &  5.32   & 1.09E9  & 1.65E16  & 7.15E4\\
16.00 &	 15837  &  7.47E3	 &  5.21   & 1.28E9  & 1.38E16  & 8.10E4\\
18.00 &	 14567  &  7.04E3	 &  5.12   & 1.47E9  & 1.17E16  & 8.95E4\\
20.00 &	 13513  &  6.67E3	 &  5.03   & 1.66E9  & 1.00E16  & 9.64E4\\
22.00 &	 12623  &  6.35E3	 &  4.96   & 1.85E9  & 8.68E15  & 1.01E5\\
26.00 &	 11196  &  5.82E3	 &  4.82   & 2.24E9  & 6.71E15  & 1.03E5\\
30.00 &	 10099  &  5.39E3	 &  4.70   & 2.59E9  & 6.58E15  & 9.59E4\\
35.00 &   9033  &  4.97E3    &  4.57   & 3.08E9  & 5.23E15  & 9.08E4\\
40.00 &   8199  &  4.62E3    &  4.45   & 3.44E9  & 1.51E14  & 8.71E4\\
45.00 &   7526  &  4.33E3    &  4.24   & 3.08E9  & 6.83E13  & 8.52E4\\
50.00 &   6970  &  4.09E3    &  4.07   & 2.82E9  & 6.97E13  & 8.31E4\\
\enddata
\tablecomments{Each line in the table represents a separate annulus.
A \citet{ss1973} viscosity parameter $\alpha=0.1$ was used in calculating all annuli.
The WD radius, $r_{\rm wd,0}$, is the radius, $0.00882R_{\odot}$, of a zero temperature 
Hamada-Salpeter carbon model. See the text (\S6.) for a discussion of the table units.
}		 
\end{deluxetable}

\clearpage
\begin{deluxetable}{rrrrrr}
\tablewidth{0pt}
\tablenum{8}
\tablecaption{Temperature profile for RW Sex 
accretion disk with mass transfer rate of
$\dot{M}=5.0{\times}10^{-9}~{M}_{\odot}{\rm yr}^{-1}$ and WD mass of $0.90{M}_{\odot}$.}
\tablehead{	  
\colhead{$r/r_{\rm wd}$} & \colhead{$T_{\rm eff}$} & \colhead{$r/r_{\rm wd}$} & \colhead{$T_{\rm eff}$} 
& \colhead{$r/r_{\rm wd}$} & \colhead{$T_{\rm eff}$}
}
\startdata
1.00    &	56744	&	20.71	&	11994	&	43.04	&	7073\\
1.14	&	56744	&	22.20	&	11413	&	44.52	&	6900\\
1.32	&	60721	&	23.69	&	10893	&	46.01	&	6736\\
2.85	&	45281	&	25.18	&	10426	&	47.50	&	6582\\
4.34	&	35070	&	26.66	&	10004	&	48.99	&	6435\\
5.83	&	28942	&	28.15	&	9620	&	50.48	&	6296\\
7.31	&	24844   &	29.64	&	9269	&	51.97	&	6164\\
8.80	&	21891	&	31.13	&	8946	&	53.46	&	6038\\
10.29	&	19651	&	32.62	&	8649	&	54.94	&	5918\\
11.78	&	17885	&	34.11	&	8374	&	56.43	&	5804\\
13.27	&	16452	&	35.59	&	8119	&	57.92	&	5695\\
14.76	&	15264	&	37.08	&	7881	&	59.41	&	5590\\
16.25   &	14260   &	38.57   &	7659	&	60.90	&	5489\\
17.73   &	13399   &	40.06   &	7451	&	62.39	&	5393\\
19.22   &	12651   &	41.55   &	7256	&	63.87	&	5301\\
\enddata
\tablecomments{
The WD radius $r_{\rm wd}$ is the radius of a 50,000K $0.90{M}_{\odot}$ WD interpolated from
Table 4a of \citet{panei2000}. The $T_{\rm eff}=56744$K for $r/r_{\rm wd}$=1.00 refers to
the inner edge of the innermost annulus. The same $T_{\rm eff}$ for $r/r_{\rm wd}$=1.14
refers to the outer edge of the innermost annulus which coincides with the inner edge of
the next annulus. The remaining $T_{\rm eff}$ values refer to the inner edge of the
corresponding annuli.
}		 
\end{deluxetable}


\begin{deluxetable}{lrrrrrrrr}
\tablewidth{0pt}
\tablenum{9}
\tablecaption{
Values of ${\chi}_{\nu}^2$
}
\tablehead{	  
\colhead{EXP} & \colhead{5.0} & \colhead{5.25} & \colhead{5.5} 
& \colhead{5.75} & \colhead{6.0} & \colhead{6.25} & \colhead{6.5}
& \colhead{7.0}
}
\startdata
0.25    &	 86.20	&	 95.31	&	103.75	&	116.87	& 130.36 & 145.71 & 159.62 & 200.61\\
0.225	&	 79.26	&	 91.19	&	102.89	&	119.26	& 136.78 & 156.84 & 174.44 & 228.35\\
0.20	&	 63.46	&	 76.56	&	 89.58	&	135.37	& 128.13 & 151.54 & 171.14 & 237.15\\
0.175	&	 39.15	&	 50.07	&	 60.87	&	 78.74	&  98.82 & 122.97 & 141.97 & 215.97\\
0.15	&	 19.74	&	 24.99	&	 30.39	&	 44.92	&  61.57 &  83.17 &  98.40 & 172.34\\
0.125	&	 23.30	&	 17.95	&    15.51  &	 21.48	&  30.55 &  44.74 &  54.40 & 115.13\\
0.10	&	 55.48  &	 38.44 	&	 31.36	&	 25.65	&  25.86 &  30.77 &  37.79 &  73.20\\
0.075	&	 95.82	&	 69.86	&	 62.42	&	 46.84	&  40.97 &  39.08 &  47.27 &  58.82\\
0.05	&	124.20	&	 93.32	&	 86.65	&	 65.77	&  57.08 &  52.00 &  62.01 &  60.47\\
0.025	&	137.22	&	107.04	&	 97.42	&	 77.25	&  67.15 &  60.60 &  67.38 &  62.87\\
0.00	&	140.81	&	112.46	&	100.65	&	 81.72	&  71.03 &  63.91 &  66.33 &  63.61\\
\enddata
\tablecomments{The second and succeeding column headings are in units of 
$10^{-9}~M_{\odot}~{\rm yr}^{-1}$. See the text for a discussion.
}		 
\end{deluxetable}


\begin{deluxetable}{lrrrrrrrr}
\tablewidth{0pt}
\tablenum{10}
\tablecaption{
Values of ${\chi}_{\nu}^2$
}
\tablehead{	  
\colhead{$T_{\rm eff}$} & \colhead{5.0} & \colhead{5.25} & \colhead{5.5} 
& \colhead{5.75} & \colhead{6.0} & \colhead{6.25} & \colhead{6.5}
& \colhead{7.0}
}
\startdata
35    &	 20.51	&	 32.49	&	 22.53	&	 18.53	&  15.68 &  20.58 &  67.42 & 137.68\\
40	&	 19.64	&	 17.19	&	 17.34	&	 26.62	&  39.25 &  60.51 &  69.78 & 141.41\\
45	&	 21.26	&	 17.36	&	 16.22	&	 23.85	&  34.71 &  52.20 &  61.89 & 128.07\\
50	&	 23.30	&	 17.95	&	 15.51	&	 21.48	&  30.55 &  44.74 &  54.40 & 115.13\\
55	&	 25.03	&	 18.64	&	 15.24	&	 19.98	&  27.71 &  39.30 &  48.99 & 105.50\\
60	&	 28.18	&	 20.19	&    15.37  &	 18.24	&  23.96 &  32.49 &  41.43 &  91.71\\
65	&	 30.88  &	 21.76 	&	 15.90	&	 17.38	&  21.61 &  27.78 &  36.19 &  81.72\\
70	&	 32.19	&	 22.62	&	 16.28	&	 17.17	&  20.68 &  25.29 &  33.82 &  76.94\\
75	&	 34.53	&	 24.19	&	 17.08	&	 16.94	&  19.32 &  22.27 &  30.20 &  69.57\\
\enddata
\tablecomments{The second and succeeding column headings are in units of 
$10^{-9}~M_{\odot}~{\rm yr}^{-1}$. The $T_{\rm eff}$ values are in 1000's of K.
See the text for a discussion.
}		 
\end{deluxetable}


\clearpage

\begin{deluxetable}{llll}
\tablewidth{0pt}
\tablenum{11}
\tablecaption{RW Sex Best Fit System Parameters}
\tablehead{
\colhead{parameter} & \colhead{value} & \colhead{parameter} & \colhead{value}}
\startdata
$M_{\rm wd}$  &  $0.90M_{\odot}$	 & $i$    &  $34{\arcdeg}$\\
$EXP$     &     0.15            &   ${M}_2$   &  $0.674{M}_{\odot}$ \\
$SCL$    &      0.51443         &  ${\dot{M}}$ & $2.0{\times}10^{-9}{M}_{\odot}~{\rm yr}^{-1}$ \\
$T_{\rm eff}$(WD) & 50,000K &   P & 0.24507 day \\
$d$ 	&$150$~pc \\	  
\enddata
\end{deluxetable}



\clearpage

\begin{figure}[tb]
\includegraphics[angle=270,scale=0.80]{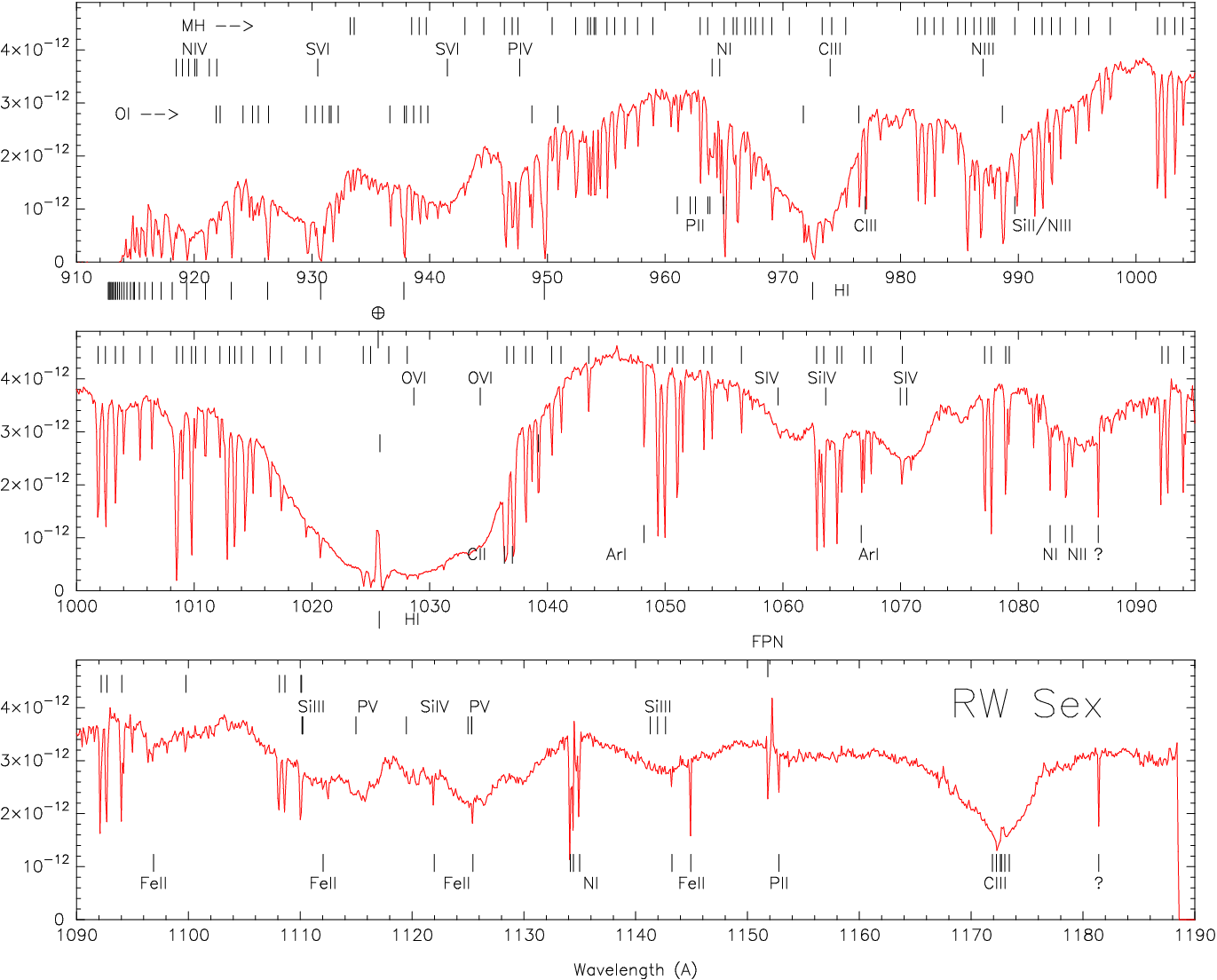}
\vspace{5pt}
\figcaption{
FUSE spectrum of RW Sex. The ordinate is flux in 
${\rm erg}~{\rm cm}^{-2}~{\rm s}^{-1}~{\rm \AA}^{-1}$
\label{fg1}}
\end{figure}

\begin{figure}[tb]
\includegraphics[angle=270,scale=0.80]{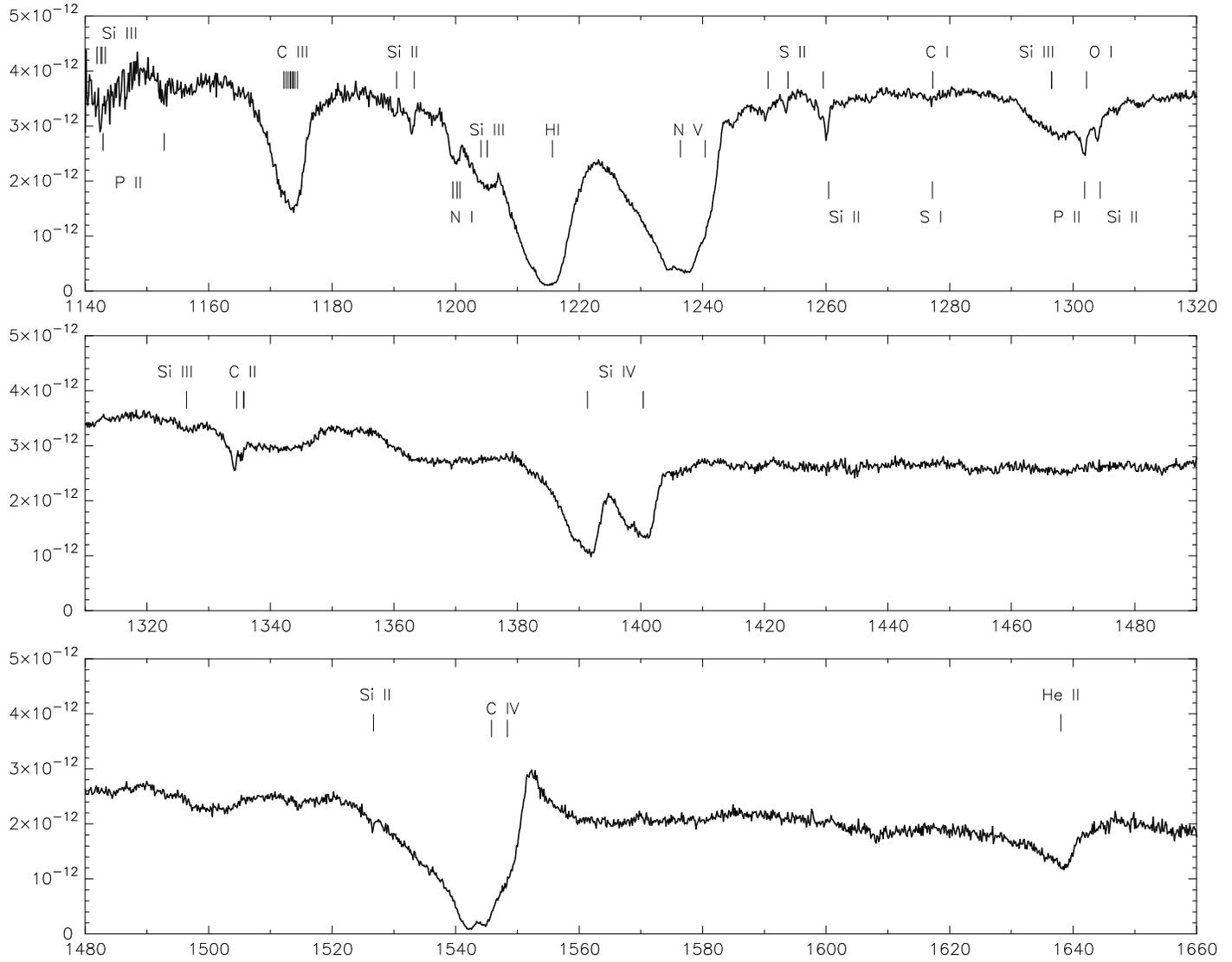}
\vspace{5pt}
\figcaption{
HST spectrum of RW Sex. The ordinate is flux in
${\rm erg}~{\rm cm}^{-2}~{\rm s}^{-1}~{\rm \AA}^{-1}$
\label{fg2}}
\end{figure}

\begin{figure}[tb]
\plotone{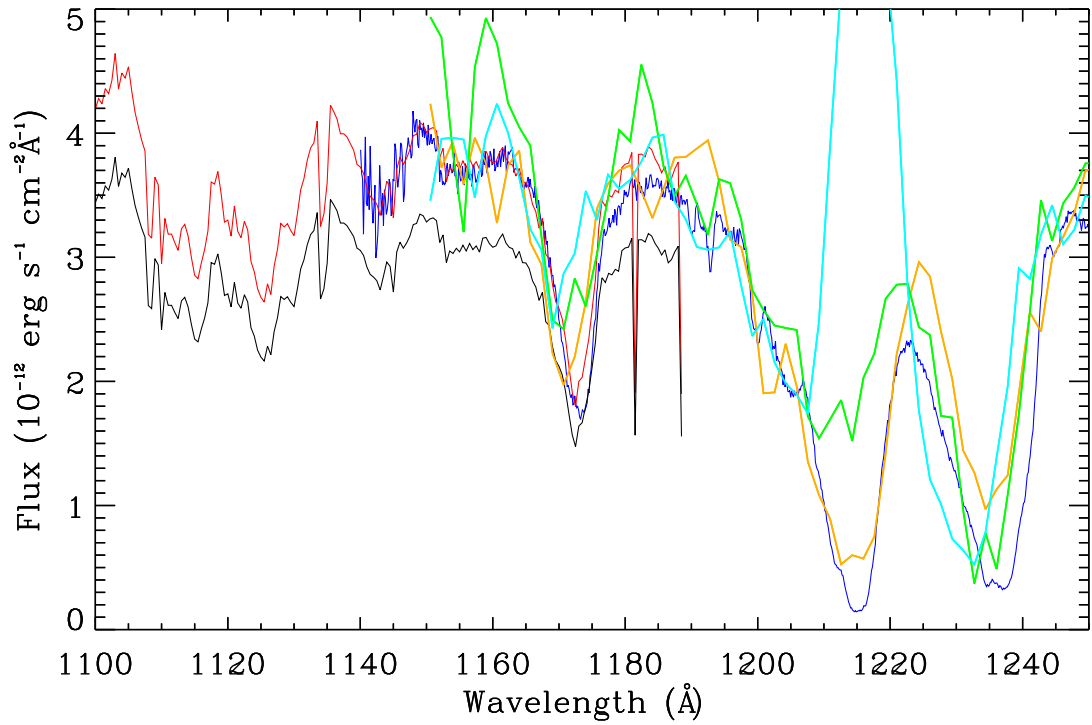}
\epsscale{0.97}
\vspace{5pt}
\figcaption{
Fit of $FUSE$, HST and $IUE$ spectra. The black line is the $FUSE$
spectrum as observed. The orange line is $IUE$ case1; the green line is
$IUE$ case2; the cyan line is $IUE$ case3. The blue line is a combination
of the mean of HST z37v0104t plus z37v0105t and the mean of
HST z37v0106t plus z37v0107t. The red line is the $FUSE$ spectrum
divided by 0.82.
See the text for a discussion.
\label{fg3}}
\end{figure}

\clearpage

\begin{figure}[tb]
\plotone{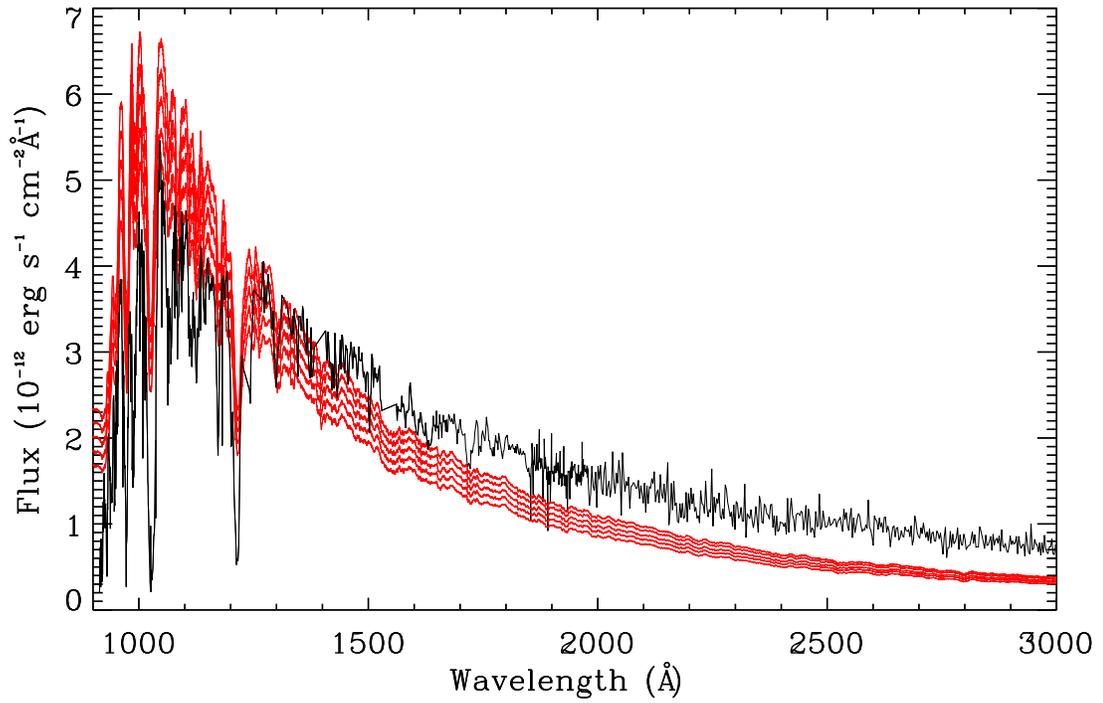}
\epsscale{0.97}
\vspace{5pt}
\figcaption{
Standard model synthetic spectrum fits to the observed (masked) spectrum of RW Sex.
The standard models are for mass transfer rates of 5.0, 5.5, 6.0, 6.5, and
$7.0{\times}10^{-9}~M_{\odot}~{\rm yr}^{-1}$ (lowest to highest flux).
\it{See the electronic edition for a color plot.}
\label{fg4}}
\end{figure}

\begin{figure}[tb]
\plotone{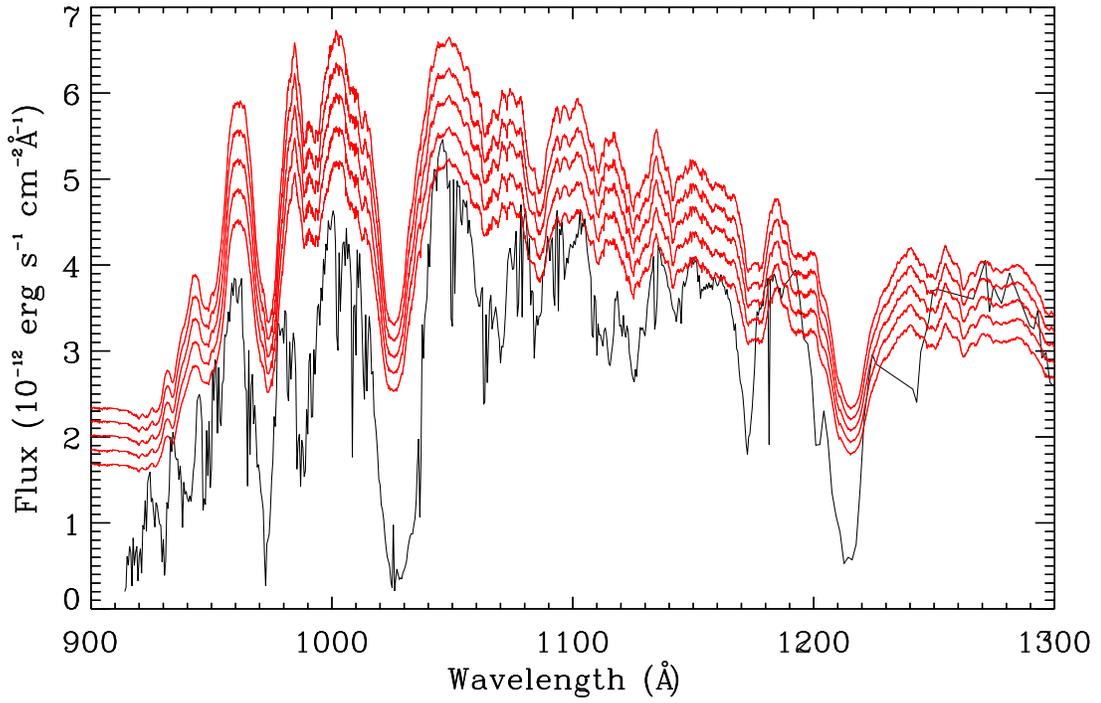}
\epsscale{0.97}
\vspace{5pt}
\figcaption{
As in Figure~4 but for the spectral region of the $FUSE$ spectrum.
\it{See the electronic edition for a color plot.}
\label{fg5}}
\end{figure}

\clearpage
\begin{figure}[tb]
\plotone{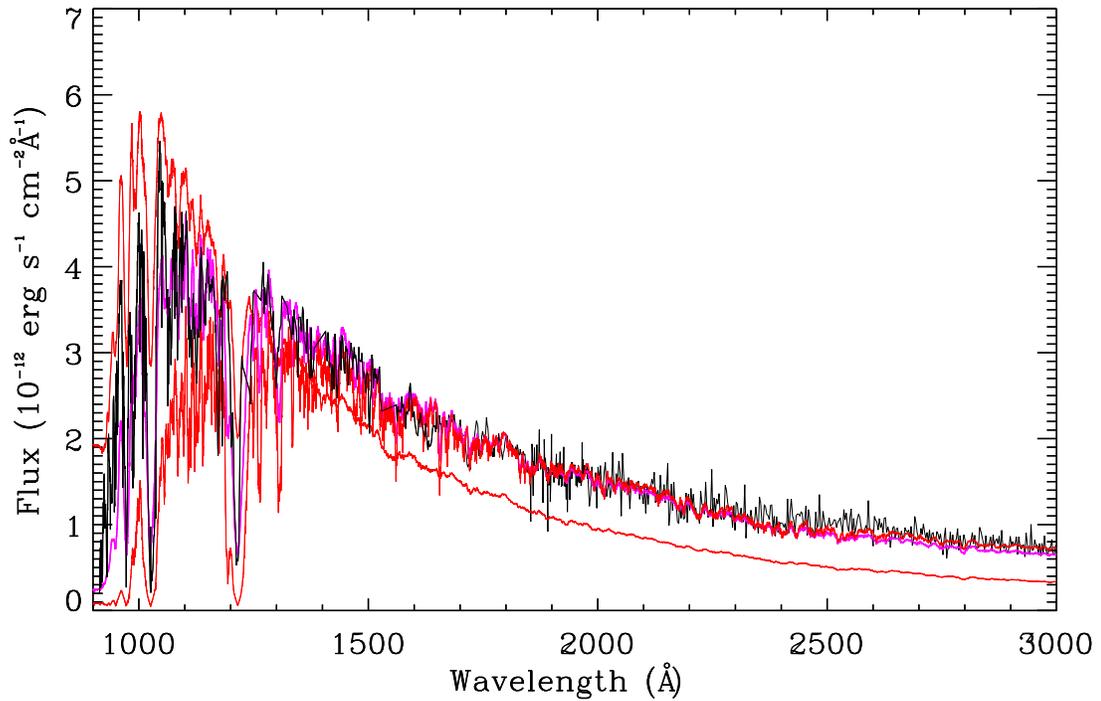}
\epsscale{0.97}
\vspace{5pt}
\figcaption{
Synthetic spectrum fits to observed data (black line). The orange line
represents the value $EXP=0.25$, the magenta line represents $EXP=0.125$,
and the red line represents $EXP=0.0$. Note that there is little difference
between the red and magenta lines at the longer wavelengths. 
\label{fg6}}
\end{figure}

\begin{figure}[tb]
\plotone{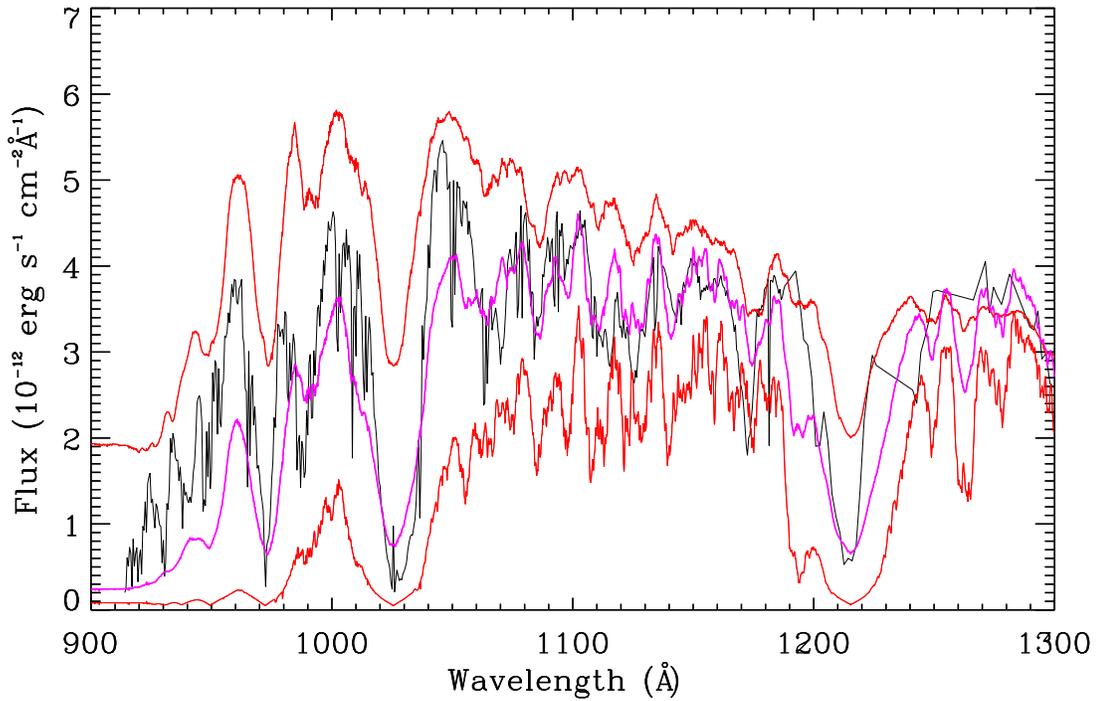}
\epsscale{0.97}
\vspace{5pt}
\figcaption{
As in Figure~6 but for a limited range in the UV. Note the orange line
 (for $EXP=0.25$) 
lies well above the observed spectrum (black line) and the Lyman lines are not deep.
The red line (for $EXP=0.0$) has broad Lyman lines and lies
well below the observed spectrum. The magenta line (for $EXP=0.125$) 
is a much better fit.
\label{fg7}}
\end{figure}

\clearpage

\begin{figure}[tb]
\plotone{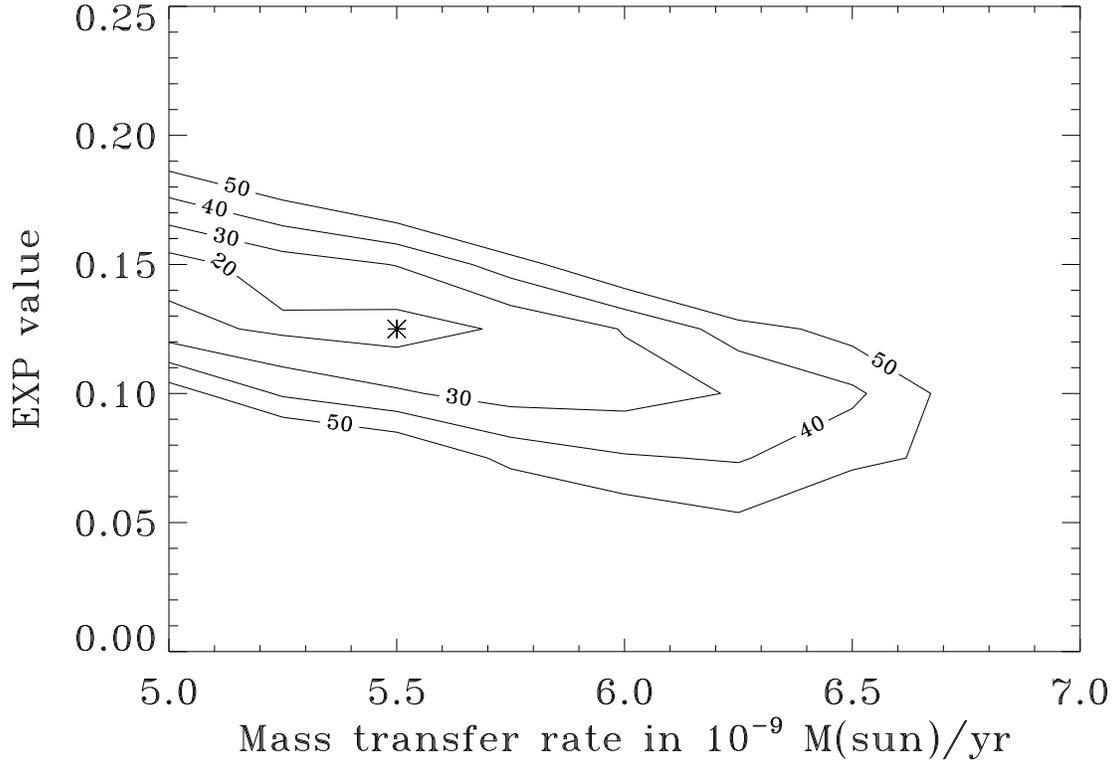}
\epsscale{0.97}
\vspace{5pt}
\figcaption{
Plot of ${\chi}_{\nu}^2$ values tabulated in Table~9. Standard model accretion disks
correspond to the top horizontal line, at EXP=0.25, bounding the plot. 
(Compare equation~2 and equation~3.)
Note the selection
of $EXP=0.125$ as roughly the best fit at all mass transfer rates. All models
adopted a WD $T_{\rm eff}=50,000$K. 
\label{fg8}}
\end{figure}

\begin{figure}[tb]
\plotone{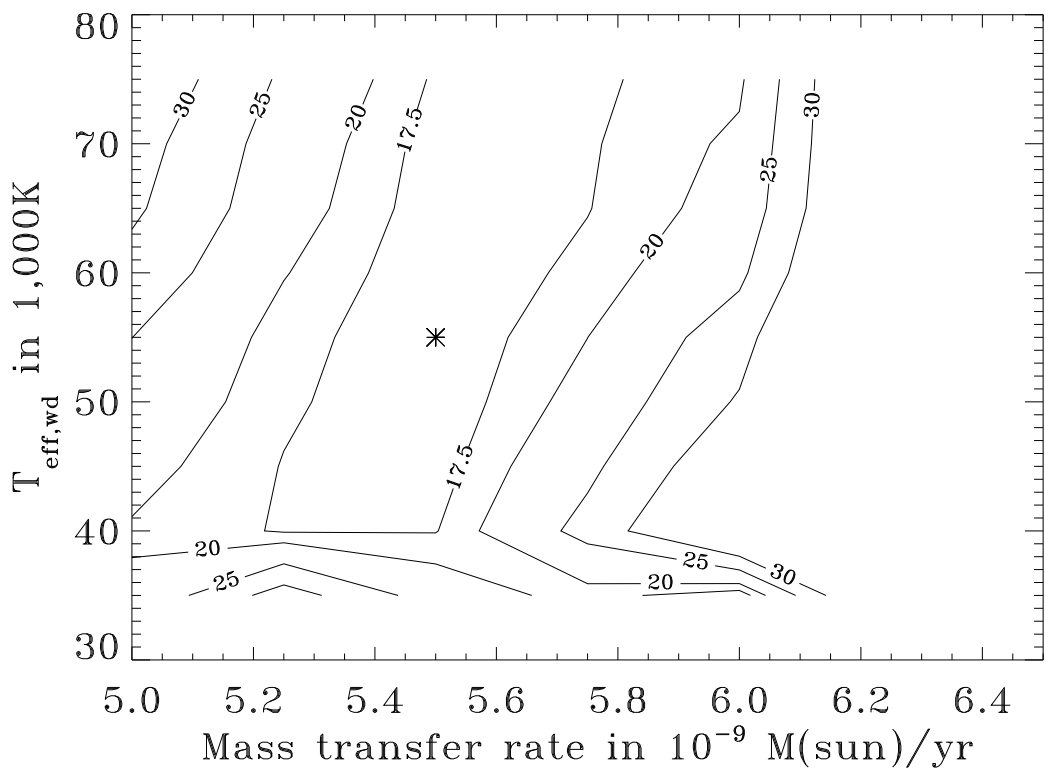}
\epsscale{0.97}
\vspace{5pt}
\figcaption{
Plot of	${\chi}_{\nu}^2$ values tabulated in Table~10. All accretion disk
models adopted a value $EXP=0.125$. The asterisk marks the poorly constrained
``best" value of $T_{\rm eff,wd}$.
As shown here, the originally adopted value
of the WD $T_{\rm eff}=50,000$K was a reasonable choice.
The lack of a closed ${\chi}_{\nu}^2$ contour limiting the $T_{\rm eff}$
range prevents determination of the WD $T_{\rm eff}$.
\label{fg9}}
\end{figure}

\begin{figure}[tb]
\plotone{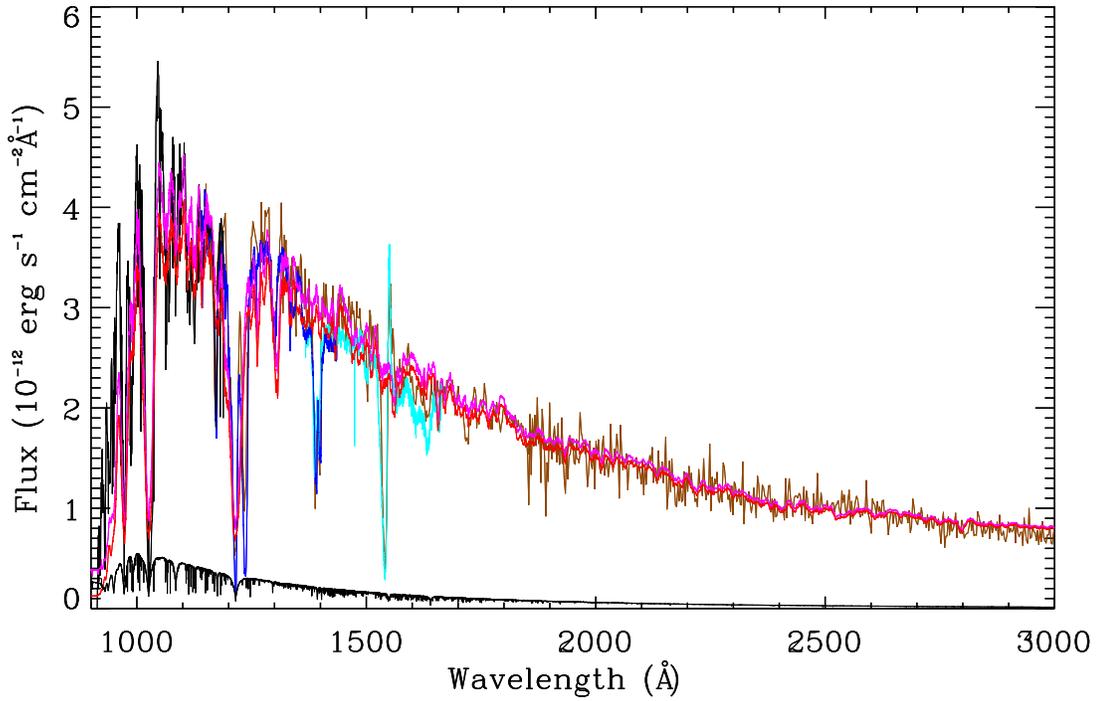}
\epsscale{0.97}
\vspace{5pt}
\figcaption{
Superposition of observed and model synthetic spectra. The upper black line is 
the $FUSE$ spectrum without masking. The brown line is the $IUE$ spectrum, without masking.
The blue line is the mean HST z37v0104t plus z37v0105t spectrum. The cyan line is the 
mean HST z37v0106t plus z37v0107t spectrum.
Note the strong chromospheric lines, which the model does not represent, that were masked in
the solution process.
The magenta line is the final model and the red line, almost indistinguishable from the
magenta line but barely below it is the accretion disk contribution. The black
line at the bottom is the 50,000K WD.
\label{fg10}}
\end{figure}

\begin{figure}[tb]
\plotone{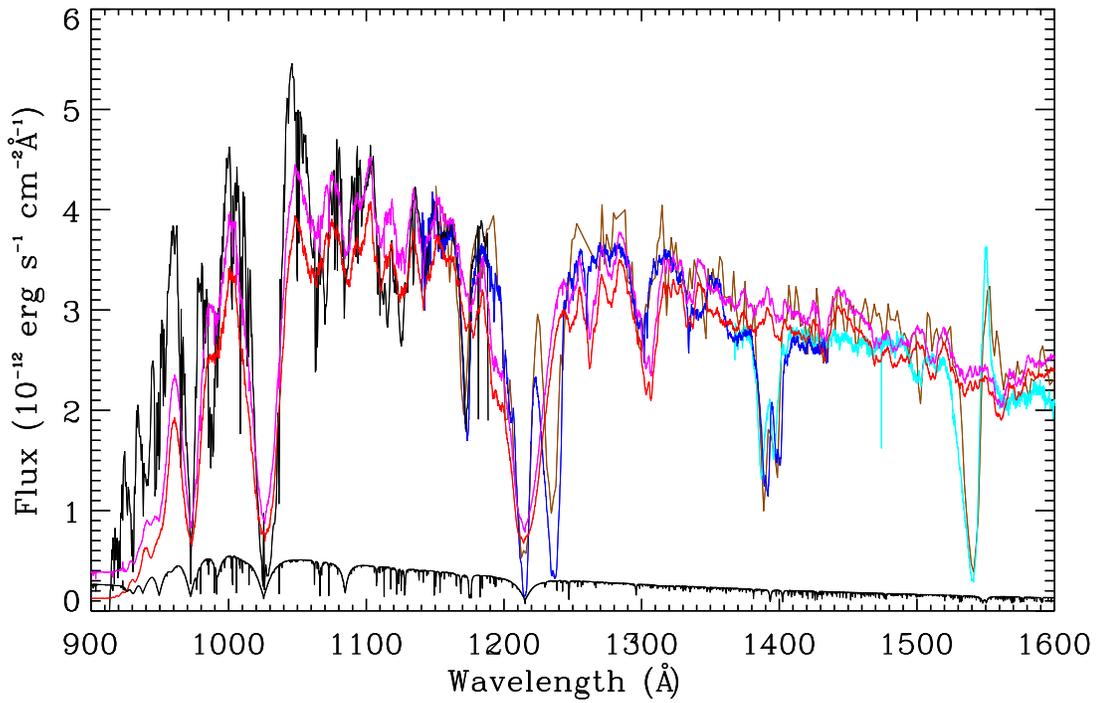}
\epsscale{0.97}
\vspace{5pt}
\figcaption{
As in Figure~10 but for a restricted wavelength range. The difference between the magenta and red
lines is barely visible. Note that this difference is the contribution of
the 50,000K WD, shown in black at the bottom of the plot.
\label{fg11}}
\end{figure}

\begin{figure}[tb]
\plotone{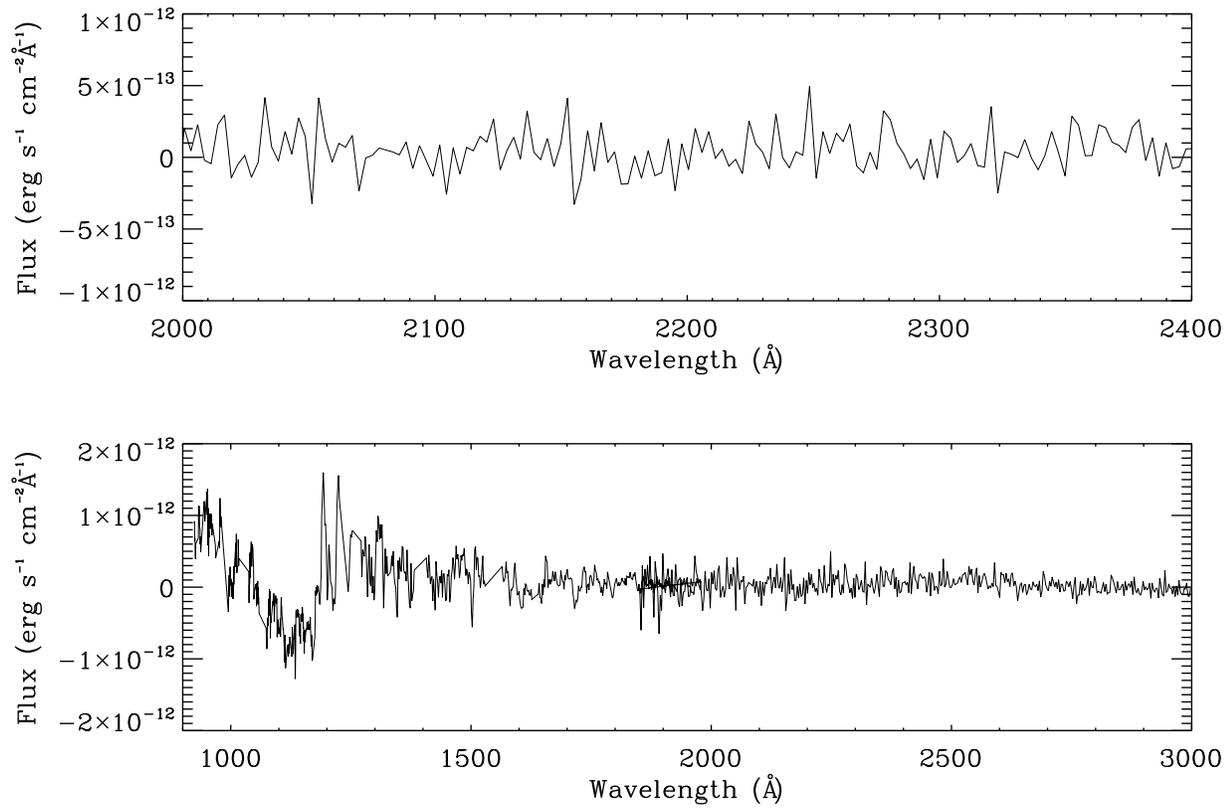}
\epsscale{0.97}
\vspace{5pt}
\figcaption{
Residuals from fit of best fit model to masked observed spectrum.
(Top) A restricted wavelength range; (Bottom) full range of combined observed spectrum.
\label{fg12}}
\end{figure}

\begin{figure}[tb]
\plotone{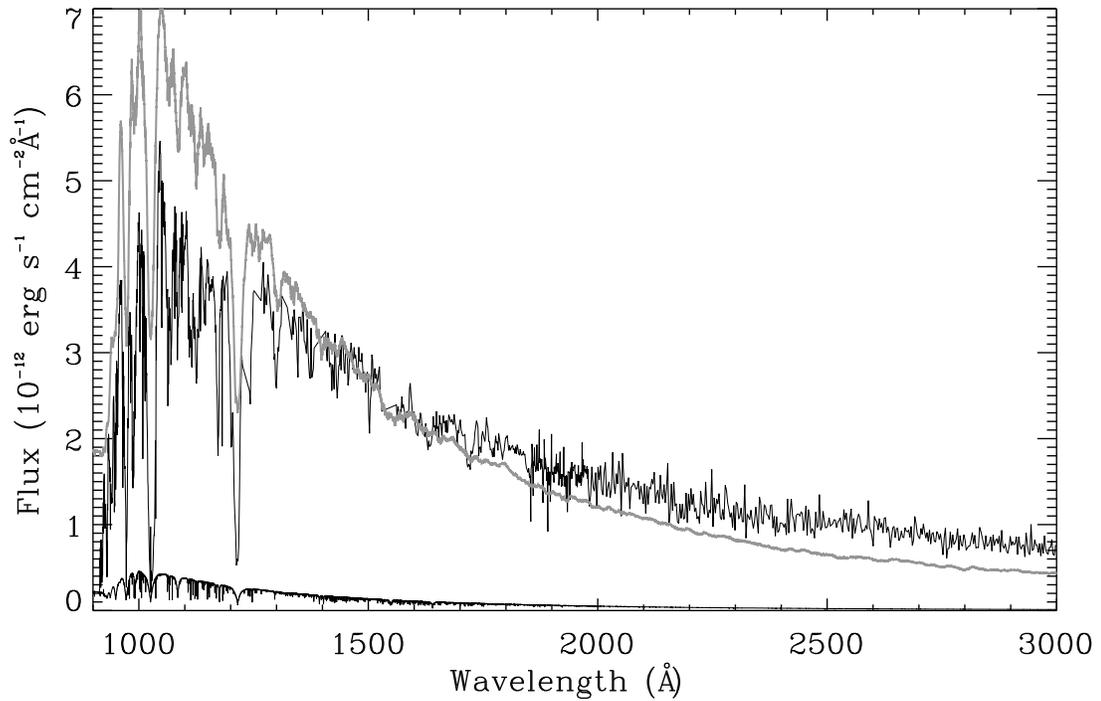}
\epsscale{0.97}
\vspace{5pt}
\figcaption{
Plot of standard model for 
$\dot{M}=2.0{\times}10^{-9}~{\rm M}_{\odot}~{\rm yr}^{-1}$ (gray line).
The observed spectrum is the masked spectrum plotted in Figure~4.
The contribution of the 50,000K WD is at the bottom.
The standard model spectrum has been divided by $3.0{\times}10^{41}$
to superpose it on the observed spectrum.
\label{fg13}}
\end{figure}

\begin{figure}[tb]
\plotone{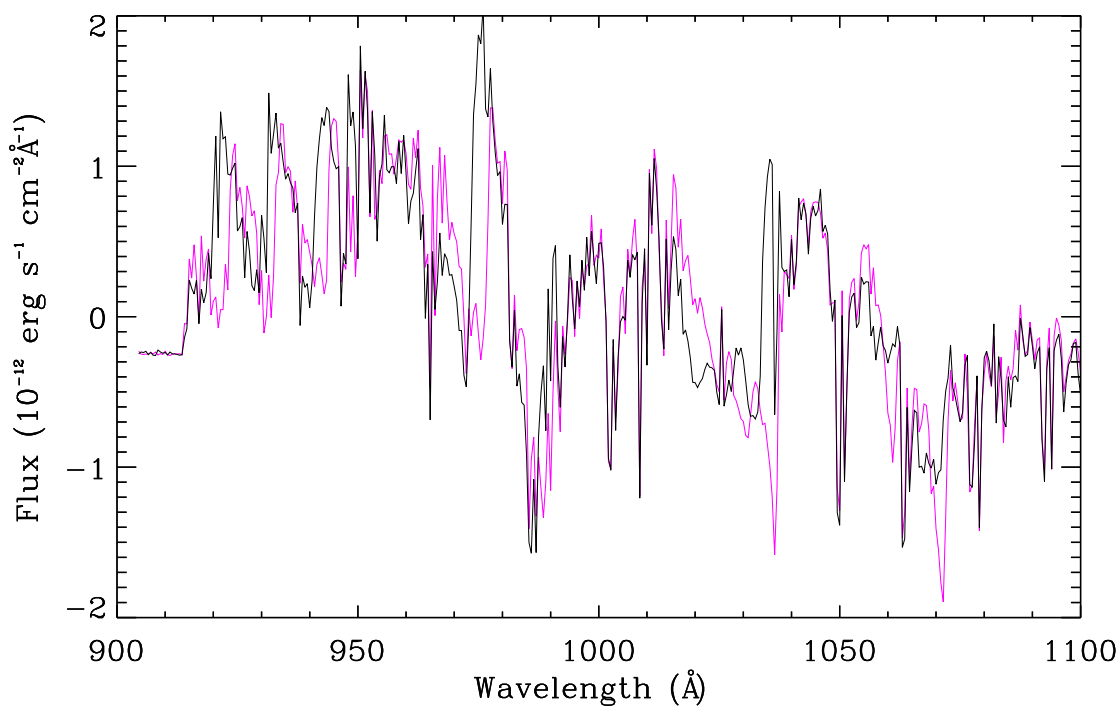}
\epsscale{0.97}
\vspace{5pt}
\figcaption{
Comparison of the second $FUSE$ time-series residuals plot (black) with
the eleventh residuals plot (magenta) at an orbital phase difference of about 0.5.
Note the large blue shift of some of the the black plot lines and the large differences
in the absorption line strengths. 
Some spectral regions match well on the two plots (950-960\AA, 990-1020\AA,
1040-1055\AA, and 1075-1100\AA), showing that there is no vertical shift of one
entire difference spectrum relative to the other.
See the text for a discussion.
\label{fg14}}
\end{figure}

\begin{figure}[tb]
\plotone{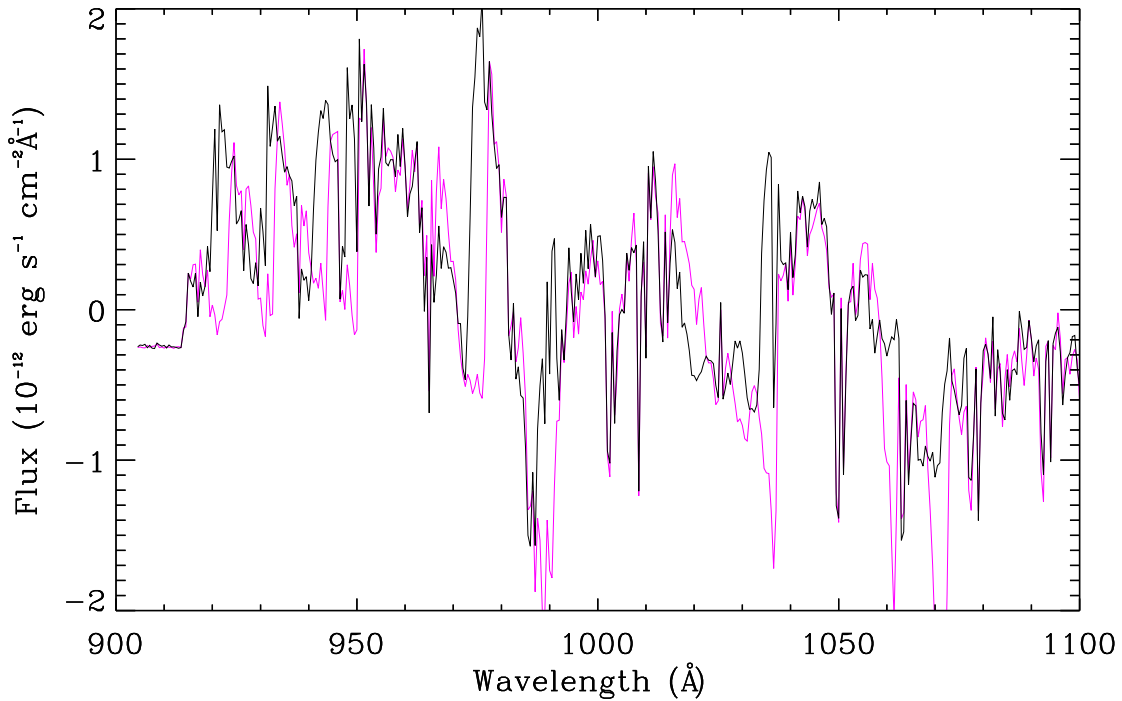}
\epsscale{0.97}
\vspace{5pt}
\figcaption{
Comparison of the second $FUSE$ time-series residuals plot (black) with
the ninth residuals plot (magenta) at an orbital phase difference of about 0.25.
Note the large differences in the absorption line strengths (black and magenta) 
and the changes from Figure~14. Also note that, as in Figure~14, the two difference
spectra match well in the spectral intervals 950-960\AA, 990-1010\AA, 1040-1055\AA,
and 1075-1100\AA. There is no overall vertical shift of one difference spectrum
relative to the other. See the text for a discussion.
\label{fg15}}
\end{figure}

\end{document}